\documentclass[fleqn,usenatbib]{mnras}

\usepackage{newtxtext,newtxmath}
\usepackage[T1]{fontenc}
\usepackage{ae,aecompl}

\usepackage{graphicx}	% Including figure files
\usepackage{amsmath}	% Advanced maths commands
\usepackage{amssymb}	% Extra maths symbols
\usepackage{soul}

\newcommand{\munu}{{\mu \nu}}
\newcommand{\del}{\partial}
\newcommand{\erfc}{\mathrm{erfc}}

\title[Modified gravity and neutrinos in the large-scale structure I: simulations]{Joint halo mass function for modified gravity and massive neutrinos I: simulations and cosmological forecasts}

\author[S. Hagstotz et al.]{
Steffen Hagstotz,$^{1,2}$\thanks{E-mail: hagstotz@usm.lmu.de}
Matteo Costanzi,$^{1,2}$
Marco Baldi$^{3,4,5}$
and Jochen Weller$^{1,2}$
\\
$^{1}$Universit\"ats-Sternwarte, Fakult\"at f\"ur Physik, Ludwig-Maximilians Universit\"at M\"unchen, Scheinerstr. 1, 81679 M\"unchen, Germany\\
$^{2}$Excellence Cluster Universe, Boltzmannstr. 2, 85748 Garching, Germany\\
$^3$Dipartimento di Fisica e Astronomia,
  Alma Mater Studiorum Universit\`{a} di Bologna, via Gobetti 93/1,
  40129 Bologna, Italy\\ $^4$Astrophysics and Space Science
  Observatory Bologna, via Gobetti 93/2, 40129, Bologna, Italy
  \\ $^5$INFN - Sezione di Bologna, viale Berti Pichat 6/2, 40127,
  Bologna, Italy
}

\date{Accepted XXX. Received YYY; in original form ZZZ}

\pubyear{2017}

\begin{document}
\label{firstpage}
\pagerange{\pageref{firstpage}--\pageref{lastpage}}
\maketitle

\begin{abstract}
We present a halo mass function accurate over the full relevant Hu-Sawicki $f(R)$ parameter space based on spherical collapse calculations and calibrated to a suite of modified gravity $N$-body simulations that include massive neutrinos. We investigate the ability of current and forthcoming galaxy cluster observations to detect deviations from general relativity while constraining the total neutrino mass and including systematic uncertainties.
%We include the effect of systematic uncertainties in the mass function and find that simple parametrised extensions such as $w$CDM are not able to detect 
% We investigate the ability of cluster abundance to constrain deviations from general relativity while varying the neutrino masses and check for the effect of theoretical systematics in the mass function.
Our results indicate that the degeneracy between massive neutrino and modify gravity effects is a limiting factor for the current searches for new gravitational physics with clusters of galaxies, but future surveys will be able to break the degeneracy.
\end{abstract}


%%%%%%%%%%%%%%%%%%%%%%%%%%%%%%%%%%%%%%%%%%%%%%%%%%

%%%%%%%%%%%%%%%%% BODY OF PAPER %%%%%%%%%%%%%%%%%%

\section{Introduction}
\label{sec:introduction}

One of the goals in modern cosmology is to understand the underlying dynamics and statistics of the cosmic density field. Clusters of galaxies trace the highest of its peaks, and theory predicts their abundance to depend exponentially on the amplitude of the matter power spectrum \citep{Press1974, Bond1991, Sheth2002} which turns them into a formidable probe of cosmological parameters \citep{Allen2011, Kravtsov2012}.

Studying the cosmic density field is especially of interest because it might reveal the mechanism for the observed accelerated expansion of the Universe. It can either be explained by introducing a smooth dark energy component to the universe's energy budget, or by modifying gravity itself. Both scenarios can potentially be tested via their imprint on the abundance of clusters \citep{Battye2003,Mohr2003}, but in this paper we will focus on the latter.

Because general relativity (GR) is the unique theory of gravity in $1+3$ dimensions under very general assumptions \citep[][]{Lovelock72}, any modifications introduce new physical degrees of freedom. While these can give rise to accelerated expansion, they also tend to enhance gravity at the perturbative level. One example discussed in this paper are the $f(R)$ scalar-tensor theories, which generalise the Einstein-Hilbert action by adding a non-linear function of the Ricci scalar $R$.

The enhancement of gravity tends to result in an increased abundance of clusters, and several approaches to model the halo mass function in modified gravity exist \citep{Kopp2013,Cataneo2016,Braun_Bates2017}. But all of these studies were performed within a one-parameter extension of the minimal $\Lambda$CDM standard model, and a natural extension is the inclusion of massive neutrinos which form a small, but unknown fraction of cosmological dark matter. The detection of a non-zero neutrino mass is firmly established by particle physics as a consequence of neutrino flavour oscillations \citep{neutrinos_Araki2005} and in cosmology the neutrino background can be measured in both the cosmic microwave background \citep{Sellentin2015} and the large scale structure \citep{Baumann2018}. Even though the mass scale is still uncertain, neutrinos lead to a suppression of structure growth below their free-streaming scale \citep{Lesgourgues2006}. This then leads to the question: Can neutrinos mask modified gravity effects in the large scale structure? Are constraints obtained on $f(R)$ theories from cluster number counts \citep{Schmidt2009,Lombriser2012, Cataneo2014} then still valid when including massive neutrinos into the analysis? And on a more fundamental level, how can the joint effects of neutrinos and modified gravity be included in the theoretical prediction of cluster abundance?

Early investigations of these issues have been presented by \citet{Baldi_etal_2014}, who performed the first $N$-body simulations of $f(R)$ gravity in the presence of massive neutrinos, clearly demonstrating a strong degeneracy between their effects on the abundance of gravitationally bound systems. More recently, \citet{Giocoli_Baldi_Moscardini_2018} and \citet{Peel_etal_2018} explored the same degeneracies based on a combination of cluster counts and weak lensing statistics along the past light cone. In this work, we continue investigating the combined effects of $f(R)$ and massive neutrinos by developing a theoretical model of the joint halo mass function, calibrated to a suite of specifically designed $N$-body simulations.

We start with a brief summary of $f(R)$ gravity in Sec.~\ref{sec:f_R_review} and present the simulation suite used to explore joint effects of modified gravity and neutrinos in Sec.~\ref{sec:simulations}. In Sec.~\ref{sec:mass_function} we introduce the joint mass function and apply our framework to forecast the ability of current and future surveys to constrain $f(R)$ theories in Sec.~\ref{sec:forecasts}. We summarise our results in Sec.~\ref{sec:conclusions}.

\section{Review of $f(R)$ gravity}
\label{sec:f_R_review}

We start from the modified Einstein-Hilbert action\footnote{We use natural units $c = \hbar = 1$}
\begin{equation}
\label{eq:Einstein-Hilbert-action}
S = \int \mathrm d x^4 \sqrt{-g} \left( \frac{R + f(R)}{16 \pi G} + \mathcal{L}_m \right) \: ,
\end{equation}
with the Lagrangian of the matter fields $\mathcal{L}_m$. We adopt the functional form proposed by \cite{Hu2007}
\begin{equation}
f(R) = - 2 \Lambda \frac{R}{R + m^2} \: ,
\end{equation}
with a constant $\Lambda$ and the curvature scale $m^2$. Note that $f(R) \rightarrow 0$ for $R \rightarrow 0$, in that sense the model does not contain a cosmological constant. For $m^2 \ll R$, the function can be expanded to get
\begin{equation}
\label{eq:f_R}
f(R) \approx -2 \Lambda - f_{R0} \frac{\bar R_0^2}{R} \: ,
\end{equation}
where $\bar R_0$ is the Ricci scalar today, overbars denote background quantities and we introduced the dimensionless parameter $f_{R0} \equiv - 2 \Lambda m^2 / \bar R_0^2$. To recover the well-measured $\Lambda$CDM expansion history, we fix the first term to the cosmological constant in GR $\Lambda = \Lambda_\mathrm{GR}$ and $f_{R0} \ll 1$ is the only remaining free parameter of the model. This implies that background quantities are indistinguishable from $\Lambda$CDM.

The modified Einstein equations are obtained by variation of Eq.~\ref{eq:Einstein-Hilbert-action} with respect to the metric $g_{\mu \nu}$
\begin{equation}
\label{eq:modified_einstein}
G_\munu - f_R R_\munu - \left( \frac{f}{2} - \Box f_R \right) g_\munu - \nabla_\mu \nabla_\nu f_R = 8 \pi G T_\munu \: ,
\end{equation}
with the new scalar degree of freedom $f_R \equiv \mathrm d f / \mathrm d R$. The trace of Eq.~\ref{eq:modified_einstein} leads to an equation of motion for the scalar field $f_R$
\begin{equation}
\label{eq:field_equation_fR}
\nabla^2 \delta f_R = \frac{a^2}{3} \big( \delta R(f_R) - 8 \pi G \delta \rho_m \big) \: ,
\end{equation}
where we adopted the quasi-static approximation and consider small perturbations on a smooth background, i.e. the quantities $\delta x \equiv x - \bar x$. The time-time component of the modified Einstein equations gives a Poisson-like equation for the scalar metric perturbation $2 \psi = \delta g_{00} / g_{00}$
\begin{equation}
\label{eq:Poisson_fR}
\nabla^2 \psi = \frac{16 \pi G}{3} a^2 \rho_m - \frac{a^2}{6} \delta R(f_R) \: ,
\end{equation}
which can still be identified with the Newtonian potential but has contributions from both the matter density $\rho_m$ and the scalar field via $\delta R(f_R)$. Eqs.~\ref{eq:field_equation_fR} and \ref{eq:Poisson_fR} are non-linear and thus we will later resort to $N$-body simulations to solve them in general, but two limiting cases are insightful:

For large field values $|f_{R0}| \gg | \psi |$ we can linearise
\begin{equation}
\delta R \simeq \left .\frac{\mathrm d R}{\mathrm d f_R} \right \rvert_{R = \bar R} \delta f_R \: ,
\end{equation}
and the Fourier-space solution of Eqs.~\ref{eq:field_equation_fR} and \ref{eq:Poisson_fR} becomes
\begin{equation}
\label{eq:Poisson_large_field}
k^2 \psi(k) = -4 \pi G \left( \frac{4}{3} - \frac{1}{3} \frac{\mu^2 a^2}{k^2 + \mu^2 a^2} \right) a^2 \delta \rho_m(k) \: ,
\end{equation}
where we introduced the Compton wavelength of the scalar field $\mu^{-1} = (3 \mathrm d f_R / \mathrm d R)^{1/2}$. On small scales $k > \mu$ this leads to a Poisson equation with an additional factor $4/3$. For scales larger than the Compton wavelength the additional contribution vanishes and we recover behaviour as in general relativity.

In the opposite limit of small field values $| f_{R0} | \ll | \psi | $ the two contributions in Eq.~\ref{eq:field_equation_fR} approximately cancel, therefore
\begin{equation}
\delta R \approx 8 \pi G \delta \rho_m
\end{equation}
and Eq.~\ref{eq:Poisson_fR} turns into the usual Poisson equation. This is the \textit{screened regime}.

To estimate where the transition occurs, we can formally solve Eq.~\ref{eq:field_equation_fR} using the Greens's function of the Laplacian
\begin{align}
\label{eq:f_R_solution}
\delta f_R(r) &= \frac{1}{4 \pi r} \frac{1}{3} \int_0^r \mathrm d^3 \mathbf{r^\prime} 8 \pi G \left( \delta \rho - \frac{\delta R}{8 \pi G} \right) \\
 &= \frac{2}{3} \frac{G M_\mathrm{eff}(r)}{r}
\end{align}
with an effective mass $M_\mathrm{eff}$ as the source for field fluctuations $\delta f_R$ \citep{Schmidt_2010}. Note that $M_\mathrm{eff}(r) \leq M(r)$ and equality holds in the unscreened regime where we get $\delta f_R = \frac{2}{3} \psi_N$ with the Newtonial potential of a spherical overdensity $\psi_N = GM / r$. Because the fluctuation in $f_R$ is by definition smaller than its background value $\delta f_R \leq \bar{f_R}$, this translates to
\begin{equation}
\label{eq:thin_shell}
| f_{R}| \leq \frac{2}{3} \psi_N(r) \: ,
\end{equation}
thus the additional force is only sourced by mass outside of the radius where this condition is met.

To summarise, the theory is identical to $\Lambda$CDM on the background level, but perturbatively yields a maximum enhancement of gravity by $1/3$ on scales smaller than the Compton wavelength $\mu^{-1}$. It also includes a screening mechanism that restores GR in regions of high density and its onset is given by the typical depth of cosmological potential wells $\psi \sim 10^{-5} - 10^{-6}$, so that $| f_{R0} |\sim 10^{-5} - 10^{-6}$ is the relevant parameter space where this mechanism can function. Values of $f_{R0}$ below this threshold are always screened, and therefore phenomenologically uninteresting.

\section{The DUSTGRAIN-{\em pathfinder} simulations}
\label{sec:simulations}

For our analysis we make use of the halo catalogues extracted from the {\small DUSTGRAIN}-{\em pathfinder} simulations \citep[see][for a detailed description]{Giocoli_Baldi_Moscardini_2018}, a suite of cosmological N-body simulations designed to investigate the possible observational degeneracies between $f(R)$ gravity and massive neutrinos by sampling their joint parameter space. The simulations have a periodic box size of $750$ Mpc$/h$ per side filled with $768^{3}$ dark matter particles of mass $m^p_{\rm cdm}= 8.1\times 10^{10}$ M$_{\odot }/h$ (for the case of $m_{\nu }=0$) and with as many neutrino particles (for the case of $m_{\nu }>0$). The particles are moving under the effect of an $f(R)$ gravitational interaction mediated by the scalar potential $\psi $ satisfying Eq.~\ref{eq:Poisson_fR} above.

The {\small DUSTGRAIN}-{\em pathfinder} runs have been performed with the \texttt{MG-Gadget} code \citep[][]{Puchwein_Baldi_Springel_2013} -- a modified version of the {\small GADGET} code \citep{gadget-2} for $f(R)$ gravity theories -- combined with the particle-based implementation of massive neutrinos developed by \citet{Viel_Haehnelt_Springel_2010}, and already employed in \citet{Baldi_etal_2014}.
The \texttt{MG-Gadget} $f(R)$ solver has been thoroughly tested \citep[see e.g.][]{Winther_etal_2015} and already used for several applications in cosmology ranging from pure collisionless simulations \citep[][]{Baldi_Villaescusa-Navarro_2018,Arnold_etal_2018} to hydrodynamical simulations \citep[][]{Arnold_Puchwein_Springel_2015,Roncarelli_Baldi_Villaescusa-Navarro_2018}, to zoomed simulations of Milky Way-sized objects \citep[][]{Arnold_Springel_Puchwein_2016,Naik_etal_2018}.

Initial conditions have been produced by generating two separate but fully correlated random realisations of the linear density power spectrum for CDM and massive neutrino particles as computed by the Einstein-Boltzmann code {\small CAMB} \citep[][]{CAMB} at the starting redshift of the simulation $z_{i}=99$. Following the approach of e.g. \citet{Zennaro_etal_2017,Villaescusa-Navarro_etal_2018}, neutrino gravitational velocities are calculated based on the scale-dependent growth rate $D(z_{i},k)$ for the neutrino component. On top of these, neutrino particles also receive an additional thermal velocity extracted from the neutrino momentum distribution for each value of neutrino mass under consideration.

In the present work -- which is the third in a series of papers making use of the {\small DUSTGRAIN}-{\em pathfinder} simulations after \citet{Giocoli_Baldi_Moscardini_2018} and \citet{Peel_etal_2018} -- we restrict our focus on a subset of the full simulations suite consisting of nine runs whose parameters are summarised in Table \ref{tab:sims}. All simulations share the same standard cosmological parameters which are set in accordance with the Planck 2015 constraints \citep[][]{Planck_2015_XIII}, namely $\Omega _m=\Omega _{\rm cdm}+\Omega _{ b}+\Omega _{\nu} = 0.31345$, $\Omega _{ b}= 0.0481$, $\Omega _{\Lambda }= 0.68655$, $H_{0}= 67.31$ km s$^{-1}$ Mpc$^{-1}$, ${\cal{A}} _{\rm s}= 2.199\times 10^{-9}$, $n_{s}=0.9658$.

\begin{table*}
\begin{tabular}{lccccccc}
Simulation Name & Gravity type  &  
$|f_{R0}|$ &
$\sum m_{\nu }$ $[\mathrm{eV}]$ &
$\Omega _{\rm cdm}$ &
$\Omega _{\nu }$ &
$m^{p}_{\rm cdm}$ $[M_{\odot }/h]$ &
$m^{p}_{\nu }$ $[M_{\odot }/h]$ \\
\hline \hline
$\Lambda $CDM & GR & -- & -- & 0.31345 & -- & $8.1\times 10^{10}$  & -- \\
fR4 & $f(R)$  & $ 10^{-4}$ & -- & 0.31345 & -- & $8.1\times 10^{10}$  & --\\
fR5 & $f(R)$  & $ 10^{-5}$ & -- & 0.31345 &--  & $8.1\times 10^{10}$  & --\\
fR6 & $f(R)$  & $ 10^{-6}$ & -- & 0.31345 & -- & $8.1\times 10^{10}$  & --\\
fR4-0.3eV & $f(R)$  & $ 10^{-4}$ & 0.3 & 0.30630 & 0.00715 & $7.92\times 10^{10}$ & $1.85\times 10^{9}$\\
fR5-0.15eV & $f(R)$  & $ 10^{-5}$ & 0.15 & 0.30987 & 0.00358 & $8.01\times 10^{10}$ & $9.25\times 10^{8}$ \\
fR5-0.1eV & $f(R)$  & $ 10^{-5}$ & 0.1 & 0.31107 & 0.00238 & $8.04\times 10^{10}$ & $6.16\times 10^{8}$ \\
fR6-0.1eV & $f(R)$  & $ 10^{-6}$ & 0.1 & 0.31107 & 0.00238 & $8.04\times 10^{10}$ & $6.16\times 10^{8}$ \\
fR6-0.06eV & $f(R)$  & $ 10^{-6}$ & 0.06 & 0.31202 & 0.00143 & $8.07\times 10^{10}$ & $3.7\times 10^{8}$  \\
\hline
\end{tabular}
\caption{The subset of the {\small DUSTGRAIN}-{\em pathfinder} simulations considered in this work with their specific parameters.}
\label{tab:sims}
\end{table*}

For all simulations we have identified collapsed CDM structures in each comoving snapshot by means of a Friends-of-Friends algorithm {\citep[FoF hereafter, see][]{Davis_etal_1985} on the CDM particles with linking length $\lambda = 0.16 \times d$ where $d$ is the mean inter-particle separation, retaining only structures with more than 32 particles. On top of such FoF catalogue we have run the {\small SUBFIND} algorithm \citep[][]{Springel_etal_2001} to identify gravitationally bound structures and to associate standard quantities such as the mass and the radius to the main substructure of each FoF group. The latter quantities are computed in the usual way by growing spheres of radius $R$ around the most-bound particle of each main substructure enclosing a total mass $M$ until the condition
\begin{equation}
\label{eq:M200mean}
\frac{4}{3}\pi R_{200m}^{3}\times 200 \times \Omega_m \rho _\mathrm{crit} = M_{200m}
\end{equation}
is fulfilled for $R=R_{200m}$ and $M=M_{200m}$, where $\rho _\mathrm{crit}\equiv 3H^{2}/8\pi G$ is the critical density of the universe. 

\section{joint mass function}
\label{sec:mass_function}

Dark matter halos form from collapsing regions that decouple from the background expansion. Their abundance can be related to the volume fraction of the Gaussian density field $\delta_R$ smoothed on a radius $R$ above a critical collapse threshold $\delta_c$ \citep{Press1974}. This yields the number density of halos within a mass interval $[M, M+dM]$, the halo mass function:

\begin{align}
\label{eq:dndm}
\frac{\mathrm d n}{\mathrm d M} = f(\sigma) \frac{\rho_m}{M^2} \frac{\mathrm d \ln \sigma^{-1}}{\mathrm d \ln M}
\end{align}
where $\rho_m = \Omega_m \rho_\mathrm{crit}$ is the mean density of the Universe and $f(\sigma)$ is the multiplicity function related to the collapsed volume fraction $F(M)$ occupied by halos over mass $M$ by
\begin{equation}
\label{eq:multiplicity_def}
f(\sigma) = 2 \sigma^2 \del F/ \del \sigma^2 \: .
\end{equation}
It depends on the variance of the linear density field
\begin{align}
\label{eq:density_variance}
S \equiv \sigma^2 \big(R(M),z \big)= \int \frac{ \mathrm d k}{k} \frac{k^3 P(k,z)}{2 \pi^2} W^2\big(kR(M) \big)
\end{align}
within a filter containing the mass $M = 4/3 \pi R^3 \rho_m $. The variables $M$, $R$ and $\sigma^2$ are monotonous functions of each other and can therefore be used interchangeably.

Note that even though $\sigma$ is often thought of as growing with cosmic time $\sigma(z) = D(z) \sigma_0$, in the framework of spherical collapse it is instructive to consider the threshold $\delta_c (z) = \delta_c / D(z)$ as the dynamical quantity. At early times, the density field is Gaussian and completely characterised by its variance alone. The collapse criterion is then really a criterion imposed on the \textit{initial conditions}.

If we assume a top-hat filter in Fourier space $W = \theta(k - 1/R)$, each new mode of the density field entering the filter is independent and the smoothed field performs a random walk with $R$ (or equivalently $S$) as a time variable. The problem can then be rephrased: when does a trajectory $\delta(S)$ first cross the threshold $\delta_c$ \citep{BBKS1986,Bond1991}?

Under these assumptions individual trajectories follow a Langevin equation
\begin{equation}
\frac{\del \delta}{\del S} = \eta(S) \: ,
\end{equation}
with a stochastic driving term $\eta$ defined by its mean $\langle \eta \rangle = 0$ and variance $\langle \eta(S) \eta(S') \rangle = \delta_D(S-S')$. The probability distribution $\Pi$ of trajectories then evolves according to the corresponding Fokker-Planck equation
\begin{align}
\label{eq:Fokker-Planck}
\frac{\del \Pi}{\del S} = \frac{1}{2} \frac{\del^2 \Pi}{\del \delta^2} \:,
\end{align}
with the boundary condition $\Pi(\delta,S=0)= \delta_D(\delta)$ because the Universe is homogeneous on large scales. However, trajectories can cross the barrier more than once leading to double-counting of halos. To solve this, one demands the additional boundary condition (an absorbing barrier) $\Pi(\delta=\delta_c,S)=0$.

The solution to Eq.~\ref{eq:Fokker-Planck} is then given by \citep{Bond1991}
\begin{align}
\label{eq:PS-PDF}
\Pi(\delta, \sigma^2) = \frac{1}{2 \pi \sigma^2} \left( \mathrm{e}^{-\delta^2/2\sigma^2} - \mathrm{e}^{-(2 \delta_c - \delta)^2 / 2 \sigma^2} \right) \: ,
\end{align}
where the second Gaussian term reflects the fact that trajectories end at the barrier. Omitting it lead to the missing normalisation factor 2 of the \cite{Press1974} prediction.

With the boundary condition the distribution function vanishes for $\delta > \delta_c$, so we express $F(S)$ by subtracting the fraction of trajectories that did not yet cross the threshold
\begin{align}
F(\sigma^2) = 1-\int_{-\infty}^{\delta_c} \Pi(\delta,\sigma^2) d \delta \: ,
\end{align}
from which we can derive the multiplicity function $f(\sigma)$ by using Eq.~\ref{eq:multiplicity_def} to get the mass function by \cite{Press1974}
\begin{align}
f_k(\sigma) = \sqrt{\frac{2}{\pi}} \frac{\delta_c}{\sigma} \mathrm{e}^{-\delta_c^2/(2\sigma^2)}  \: ,
\end{align}
with the correct normalisation. Note that we indicate solutions from non-correlated random walks (using a $k$-space top-hat) with subscript $k$.

This approach works reasonably well, but has several shortcomings:
\begin{enumerate}
\item Collapse in a Gaussian random field does not occur spherically. In the Zel'dovich approximation, the eigenvalues $\lambda_i$ of the deformation tensor follow the joint probability distribution \citep{Doroshkevich1970}
\begin{align}
p(\lambda_1, \lambda_2, \lambda_3) =& \frac{15^3}{8 \pi \sqrt{5} \sigma^6} \exp \left( - \frac{3 I_1^2}{ \sigma^2} + \frac{15 I_2}{2 \sigma^2}\right) \\ 
& \times  | (\lambda_3 - \lambda_2) (\lambda_3 - \lambda_1) (\lambda_2 - \lambda_1) | \: ,
\end{align}
with $I_1 = \lambda_1 + \lambda_2 + \lambda_3$ and $ I_2 = (\lambda_1 \lambda_2 + \lambda_1 \lambda_3 + \lambda_2 \lambda_3)$. Isotropic collapse with $\lambda_1 = \lambda_2 = \lambda_3$ therefore does not occur. Instead the Zel'dovich picture suggests a collapse into subsequently walls, sheets, filaments and halos, where the last step occurs typically along a filament in an ellipsoidal fashion. This is fully consistent with structure formation observed in $N$-body simulations.
\item Real halos do not form out of sharp $k$-space top-hats. Usually one assumes rather a real-space top-hat as initial condition for the spherical collapse. This leads to coupling of Fourier modes and introduces correlations between steps of the random walk.

\end{enumerate}

\subsection{Diffusing, drifting barrier}

\begin{figure}
	\includegraphics[width=\columnwidth]{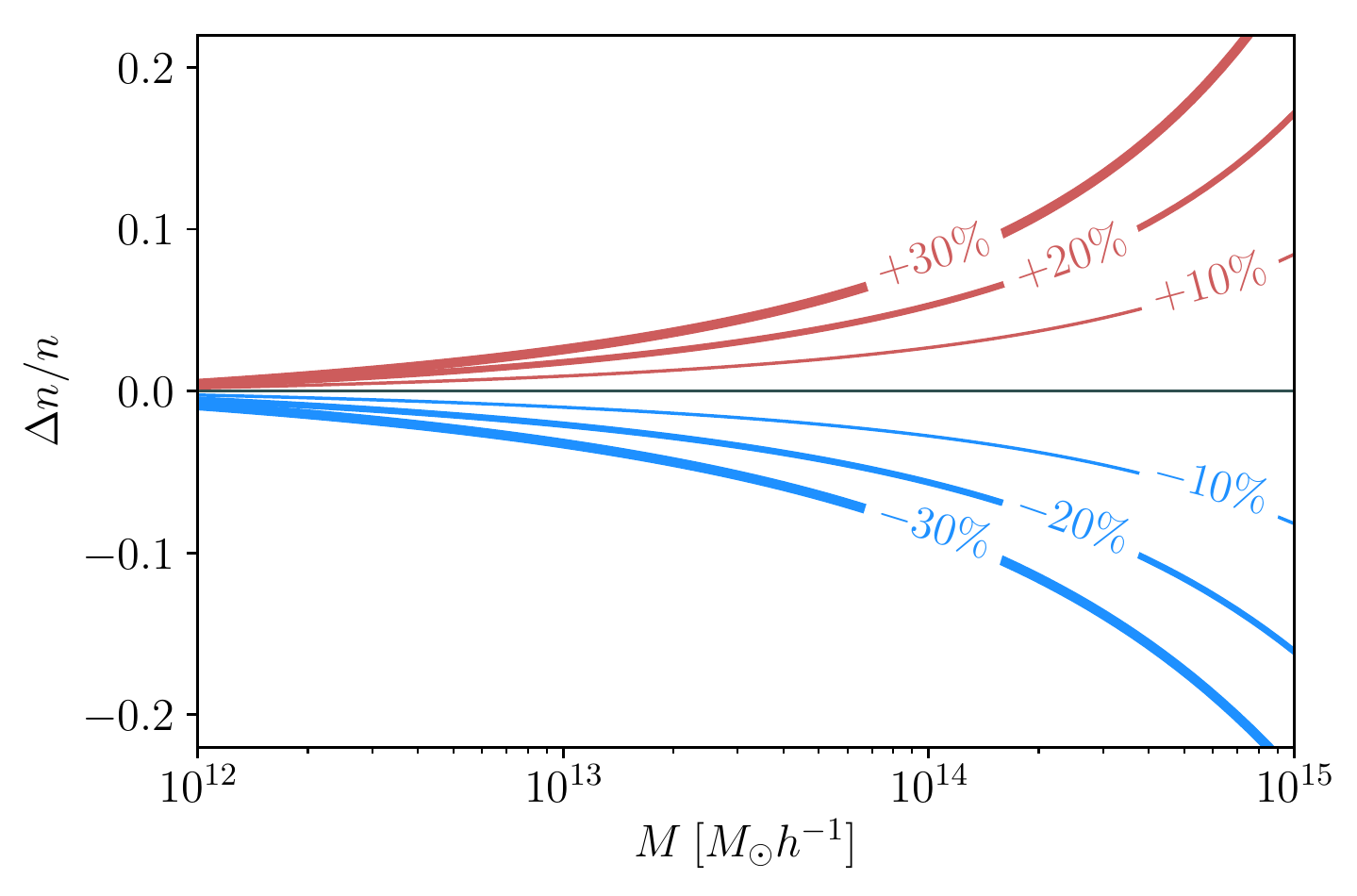}
    \caption{Effect on the halo mass function when changing the barrier width $D_B$ around the fiducial value $D_B = 0.4$. A broader barrier leads to a smaller suppression of the exponential tail of the mass function and therefore more high-mass objects.}
    \label{fig:barrier_DB}
\end{figure}

The non-spherical collapse dynamics can be addressed by modifying the collapse barrier. The main motivation is that low-mass (high $\sigma$) halos are more ellipsoidal, while the largest objects are approximately spherical. Ellipsoidal patches collapse later because they have to get rid of angular momentum, which leads to an effective higher threshold. There are various ways to extend the excursion set formalism to account for this, and here we follow \cite{Kopp2013} and introduce a scale-dependent barrier of the form
\begin{align}
\label{eq:drifting_barrier}
B = \delta_c + \beta S \: ,
\end{align}
that tends to the spherical collapse threshold $\delta_c$ for high-mass halos $\sigma \ll 1$.
Even though more general forms for the ellipsoidal collapse barrier $B$ can be found in the literature \citep[e.g. $B = \delta_c + \beta^\gamma S$; see][]{Sheth2002}, the linear approximation adopted in this work is sufficient for typical cluster abundance studies using clusters of mass $M \gtrsim 10^{13.5} M_\odot / h$.

In addition to the barrier drift, the collapse dynamics themselves are complicated by environmental effects and fuzzy halo definitions. In \cite{MaggioreII} this was taken into account by turning the barrier itself into a Gaussian stochastic variable with a mean $\bar B=\delta_c + \beta S$ and width $D_B$. Both the trajectories and the barrier itself perform a random walk, and the  joint probability distribution is obtained from a 2D Fokker-Planck equation \citep{MaggioreII, Corasaniti2011}
\begin{equation}
\frac{\del \Pi}{\del S} = \frac{1}{2} \frac{\del^2 \Pi}{\del \delta^2} + \frac{D_B}{2} \frac{\del^2 \Pi}{\del B^2} \:,
\end{equation}
which is solved by
\begin{align}
\label{eq:multiplicity_general_barrier}
f_k(\sigma) = \sqrt{\frac{2a}{\pi}} \frac{1}{\sigma} \mathrm{e}^{-a \bar B^2/(2 \sigma^2)} \left( \bar B - \sigma^2 \frac{\mathrm{d} \bar B}{\mathrm{d} \sigma^2} \right) \: ,
\end{align}
with $a \equiv 1/(1+D_B)$. Using Eq.~\ref{eq:drifting_barrier}, this reduces to a Press-Schechter like solution with the constant threshold $\delta_c$ replaced by the full barrier:
\begin{equation}
\label{eq:f_k_GR}
f_k(\sigma) = \sqrt{\frac{2 a}{\pi}} \frac{\delta_c}{\sigma} \mathrm{e}^{-a(\delta_c + \beta \sigma^2)^2 / 2 \sigma^2} \: .
\end{equation}

The effect of $D_B$ is demonstrated in Fig.~\ref{fig:barrier_DB}: a broader barrier leads to a smaller factor $a$ in the exponential, boosting the abundance of high-mass clusters because those rare trajectories can cross the threshold easier.

\subsection{Non-Markovian Corrections}

Accounting for realistic filter functions makes it necessary to consider the deviations from an uncorrelated random walk. Halos form from regions that resemble spherical patches in the initial conditions and several possible window functions to capture the correct form of these proto-halos exist \citep{Bond1991}. Here we assume a real space top-hat, which in Fourier space turns into
\begin{equation}
W(x) = \frac{3 j_1(x)}{x} \: ,
\end{equation}
with the spherical Bessel function $j_n$, which we use from here on to calculate the variance of the density field $S$ in Eq.~\ref{eq:density_variance}. In \cite{MaggioreI} the authors calculated the corrections induced by correlations between the variance $S$ smoothed at different radii $R$ for this choice of smoothing filter. The general two-point correlation function can be written as 
\begin{align}
\langle \delta_1 \delta_2 \rangle = \mathrm{min}(S_1, S_2) + \Delta(S_1, S_2) \: ,
\end{align}
where we introduced the shorthand $\delta_1 = \delta(R_1)$, and the first term expresses the Markov dynamics leading to the Press-Schechter result with a general barrier in Eq.~\ref{eq:f_k_GR}. The correction is of the form
\begin{align}
\Delta(S_1, S_2) = \kappa \frac{S_1 (S_2 - S_1)}{S_2}
\end{align}
with the coefficient
\begin{align}
\label{eq:kappa_corr}
\kappa(R) \simeq 0.459 - 0.003 R \: ,
\end{align}
and has a weak dependence on cosmology via the power spectrum. As pointed out above, we deal with a purely Gaussian field in the initial conditions here, and all correlations are introduced by the filter and not by later non-linear mode coupling. This also means that $\kappa$ should be calculated from the $\Lambda$CDM relation in Eq.~\ref{eq:kappa_corr} even within a modified gravity model. We will return to this point when discussing the modified gravity mass function.

This leads to the real-space top-hat multiplicity function $f_x$, to first order in $\kappa$ \citep{MaggioreI, Kopp2013},
\begin{align}
\label{eq:f_x_GR}
f_x(\sigma) = f_k(\sigma) + f_{1,\beta=0}^{m-m}(\sigma) + f_{\beta^{(1)}}^{m-m}(\sigma) + f_{1,\beta^{(2)}}^{m-m}(\sigma)
\end{align}
with the Markovian term $f_k$ for a diffusive, drifting barrier given by Eq.~\ref{eq:f_k_GR} and corrections
\begin{align}
f_{1,\beta=0}^{m-m}(\sigma) &= a \kappa \frac{\delta_c}{\sigma} \left( \mathrm{e}^{a \delta_c^2 / 2 \sigma^2} - \frac{1}{2} \Gamma \Big( 0, \frac{a \delta_c^2}{2 \sigma^2} \Big) \right) \: ,\\
f_{\beta^{(1)}}^{m-m}(\sigma) &= -a \delta_c \beta \left(a \kappa \: \mathrm{erfc} \Big( \delta_c \sqrt{\frac{a}{2 \sigma^2}} \Big) + f_{1, \beta = 0}^{m-m} (\sigma) \right) \: , \\
f_{1,\beta^{(2)}}^{m-m}(\sigma) &= -a \beta \left( \frac{\beta}{2} \sigma^2 f_{1,\beta=0}^{m-m}(\sigma) + \delta_c f_{1,\beta^{(1)}}^{m-m} (\sigma) \right) \: .
\end{align}

\subsection{Spherical collapse in modified gravity}

As for the $\Lambda$CDM case, the starting point of our analysis is spherical collapse. \cite{Kopp2013} numerically solved the full modified Einstein, scalar field and non-linear fluid equations to obtain $\delta_c$ in $f(R)$ gravity, and they parameterised their solution for the threshold by
\begin{align}
\delta_c^{f(R)}(f_{R0}, M, z) &= \delta_c^\mathrm{GR}(z) \times \Delta \left(f_{R0}, M, z \right) \vphantom{\Big(} \label{eq:fR_barrier}
\end{align}
where the deviation from GR is captured by the correction factor
\begin{align}
\Delta(f_{R0}, M, z) &= 1 + b_2 \left(1+z \right)^{-a_3} \left(m_b - \sqrt{m_b^2 + 1} \right) \\
&+ b_3 \Big(\tanh\left(m_b \right) - 1 \Big) \nonumber \\
m_b(f_{R0}, M, z) &= (1+z)^{a_3} \left( \log_{10} M - m_1(1+z)^{-a_4} \right) \vphantom{\Big(} \label{eq:m_b}  \\
m_1(f_{R0}) &= \mu_1 \log_{10} |f_{R0}| + \mu_2 \vphantom{\Big(} \nonumber \\
b_2 &= 0.0166 \nonumber \\
b_3(f_{R0}) &= \beta_3 \left(2.41 - \log_{10} |f_{R0}| \right) \nonumber \\ % 0.0027 \left(2.41 - \log_{10} f_{R0} \right)
a_3(f_{R0}) &= 1 + \exp \left( -2.08 \left(\log_{10} |f_{R0}| + 5.56 \right)^2 \right) \nonumber \\
a_4(f_{R0}) &= \alpha_4 \left( \tanh \Big(0.69 \left(\log_{10} |f_{R0}| + 6.65 \right) \Big) + 1 \right) \nonumber \: .
\end{align}
The parameterisation converges to the GR limit $\delta_c^\mathrm{GR}$ separately for high $z$ and $| f_{R0} | \rightarrow 0$, which is well approximated by \citep{Nakamura1997}
\begin{equation}
\label{eq:delta_c_GR}
\delta_c^\mathrm{GR}(z) = \frac{3 (12 \pi)^{2/3}}{20} \left(1 - 0.0123 \log_{10} \bigg( 1 + \frac{\Omega_m ^ {-1} - 1}{(1 + z)^3} \bigg) \right) \: .
\end{equation}
The coefficients $\alpha_4, \beta_3, \mu_1, \mu_2$ from \cite{Kopp2013} are given in Tab.~\ref{tab:fit_fiducial} which were fitted to numerical solutions and should be regarded as prediction of their spherical collapse model. Here we want to bring this model closer to data before we consider possible constraints from cluster abundance.

\begin{table}
	\centering
	\caption{Fiducial values for the GR mass function barrier shape and the virial $f(R)$ collapse threshold Eq.~\ref{eq:fR_barrier}.}
	\label{tab:fit_fiducial}
	\begin{tabular}{c c | c c c c} % six columns, alignment for each
		GR & & $f(R)$ & & & \\
        \hline \hline
		$D_B$ & $\beta$ & $\alpha_4$ & $\beta_3$ & $\mu_1$ & $\mu_2$\\
		$0.4$  & $0.12$ & $0.11$ & $2.7 \times 10^{-3}$ & $1.99$ & $26.21$ \\        
        \hline
\end{tabular}
\end{table}

\begin{figure}
	\includegraphics[width=\columnwidth]{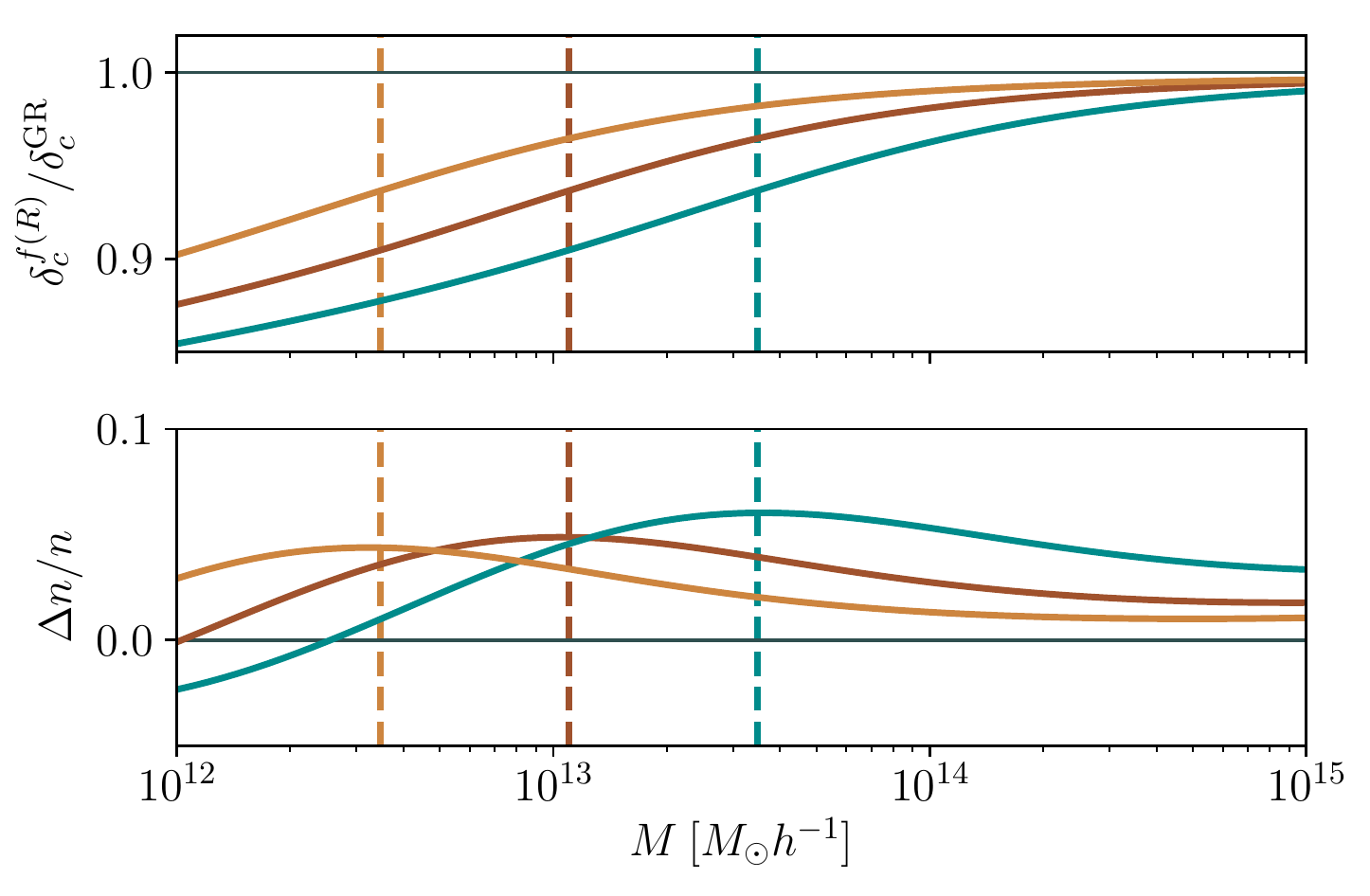}
    \caption{Top: Relative change in the collapse threshold $\delta_c / \delta_c^\mathrm{GR}$ for slightly different values of the screening mass $M_\mathrm{screen}$ (dashed vertical lines) around $f_{R0} \simeq 10^{-6}$ at $z=0$. This corresponds to the position of the typical bump in the relative cluster abundance compared to $\Lambda$CDM (bottom).}
    \label{fig:delta_c_delta_n}
\end{figure}

The crucial ingredient of the model is $m_b$, which sets the transition mass where screening sets in. We will express this scale as the screening mass $M_\mathrm{screen}$, defined by $m_b(M_\mathrm{screen}) = 0$. In Fig.~\ref{fig:delta_c_delta_n} we show the connection between the threshold and the cluster abundance: up to $M_\mathrm{screen}$ the threshold grows linearly with $\log M$ and afterwards it starts reverting to the fiducial GR value. In the mass function, this scale corresponds to a characteristic peak in the additional relative abundance. Note that the negative relative abundance for lower masses shown in the plot is physical because of mass conservation: additional high-mass objects form from low-mass halos.

For $m_b = 0$, the threshold is given by
\begin{align}
\delta_c = \delta_c^\mathrm{GR} \left( 1 + b_2 (1+z)^{-a_3} - b_3 \right) \: ,
\end{align}
and because a lower $\delta_c$ leads to a higher cluster abundance, $b_2$ and $b_3$ set the height of the additional abundance peak, $a_3$ and $a_4$ control the redshift evolution of the screening mass, and $\mu_1, \mu_2$ determine how quickly the model reverts to GR when changing $f_{R0}$.

Fig.~\ref{fig:delta_c_contour} shows the variation of the threshold as a function of redshift and the $f_{R0}$ parameter for a halo of mass $M_{200} = 10^{14} M_\odot/h$. Considering this mass representative of the lightest objects entering a cosmological cluster catalogue, the leftmost line indicates the limit of cluster abundance studies to constrain the theory at a given redshift where the deviation in $\delta_c$ is of order $1 \%$.

\begin{figure}
	\includegraphics[width=\columnwidth]{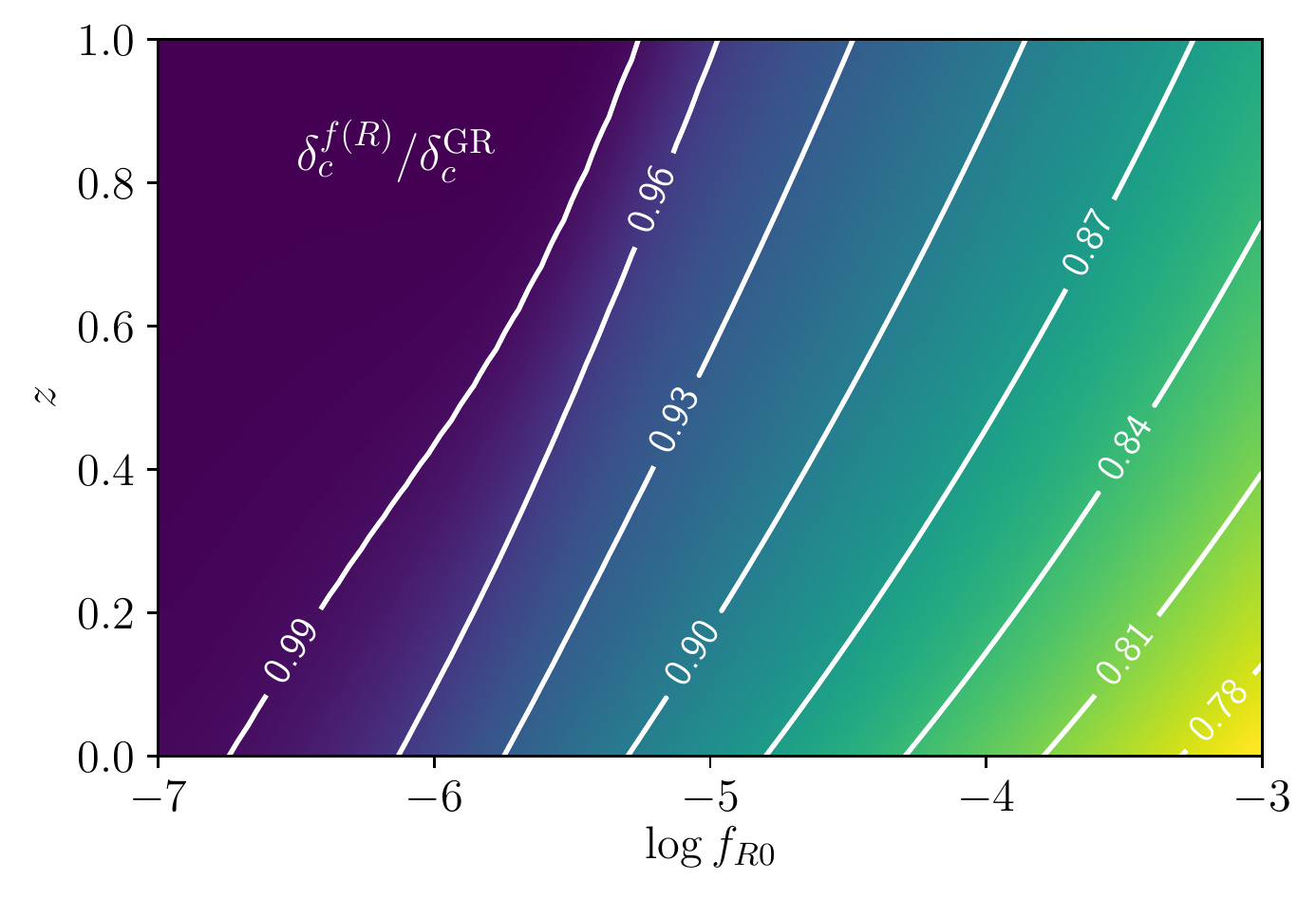}
    \caption{Change in collapse threshold $\delta_c^{f(R)} / \delta_c^\mathrm{GR}$ for a halo of fixed mass $M_{200} = 10^{14} M_\odot/h$ with redshift and $f_{R0}$. The fiducial threshold is lowered due to the fifth force for large $f_{R0}$. At high redshifts, $\delta_c$ reverts to the $\Lambda$CDM value. The plot includes the corrections from Sec.~\ref{sec:calibration}.}
    \label{fig:delta_c_contour}
\end{figure}

To write the multiplicity function for $f(R)$ including non-Markovian corrections, we assume that the correlation between steps behaves similar for modified gravity and GR. This is justified because we measure the correlation in the initial conditions where the density fields in both theories are identical -- all modifications to the time evolution are absorbed into the threshold $\delta_c(f_{R0},M,z)$. Therefore we write \citep{Kopp2013}
\begin{align}
\label{eq:f_x_fR}
f^{f(R)}_x(\sigma) \simeq f^\mathrm{GR}_x(\sigma) \frac{f^{f(R)}_k}{f^\mathrm{GR}_k}
\end{align}
with the Markovian multiplicity function $f_k^{f(R)}$ derived from the modified gravity barrier $\bar B = \delta_c(f_{R0}, M, z) + \beta \sigma^2$ given in eq.~\ref{eq:fR_barrier}
\begin{align}
\label{eq:f_k_fR}
f^{f(R)}_k(\sigma) &= \sqrt{\frac{2a}{\pi}} \frac{1}{\sigma} \mathrm{e}^{-a \bar B^2/(2 \sigma^2)} \left( \bar B - \sigma^2 \frac{\mathrm{d} \bar B}{\mathrm{d} \sigma^2} \right) \nonumber \\
&= \sqrt{\frac{2a}{\pi}} \frac{1}{\sigma} \mathrm{e}^{-a \bar B^2/(2 \sigma^2)} \left( \delta_c^{f(R)} - \frac{3 M}{2} \frac{\del \delta_c^{f(R)}}{\del M} \frac{\del \ln \sigma}{\del \ln R} \right) \: .
\end{align}
Together with $f_k^\mathrm{GR}$ (Eq.~\ref{eq:f_k_GR}) and $f_x^\mathrm{GR}$ (Eq.~\ref{eq:f_x_GR}), this defines the full modified gravity multiplicity function (Eq.~\ref{eq:f_x_fR}), and yields the halo mass function via Eq.~\ref{eq:dndm}. We emphasize again that all expressions are defined for the smoothed density field $\sigma^\mathrm{GR}$ calculated in a \textit{standard cosmology} -- as already discussed, the threshold is imposed on the initial conditions, and all subsequent effects of modified gravity are encapsulated in the dynamics of the barrier.

\subsection{Neutrinos}

As we have seen, the signal of modified gravity is a lower collapse threshold and a resulting higher abundance of clusters compared to $\Lambda$CDM. To set realistic limits on deviations from GR, we will now incorporate effects of massive neutrinos. As has been studied before \citep[see e.g.][]{Lesgourgues_Pastor_2006} they suppress structure growth below the free-streaming scale which leads to a lower abundance of galaxy clusters, counteracting possible effects of $f(R)$. Constraining the neutrino mass is an important goal for cluster cosmology in its own right, but here we will focus on degeneracy with modified gravity effects.

\begin{figure}
	\includegraphics[width=\columnwidth]{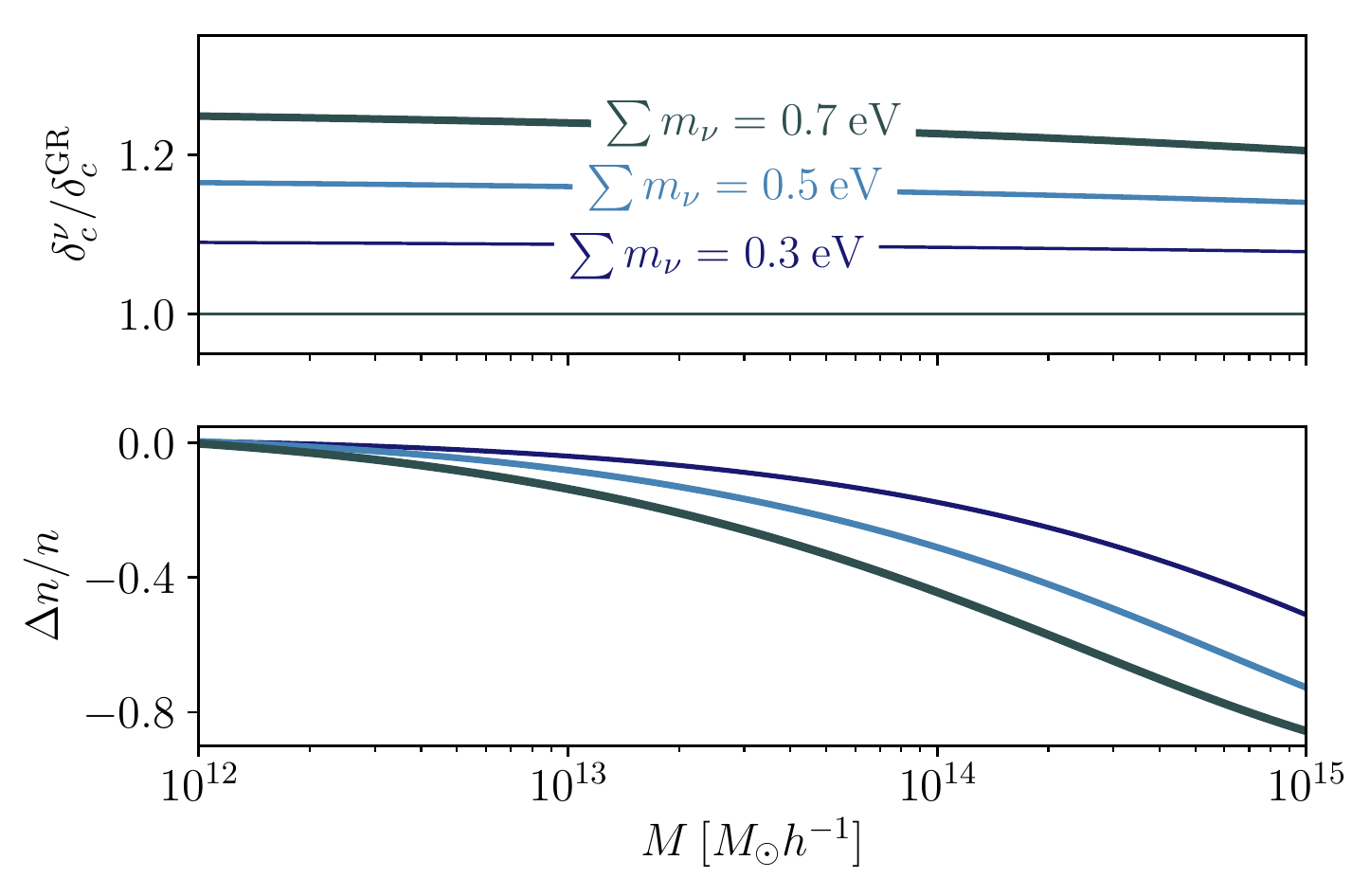}
    \caption{Top: Change in the collapse threshold $\delta_c^{\nu} / \delta_c$ for different neutrino masses. The scale dependent growth in $\nu$CDM cosmologies translates to a slight mass dependence of $\delta_c$. The higher threshold leads then to a stronger suppression in the exponential high mass tail of the mass function (bottom).}
    \label{fig:delta_c_delta_n_neutrino}
\end{figure}

\cite{Costanzi2013} showed that the effect of neutrinos on the cluster abundance can be well captured by rescaling the smoothed density field
\begin{equation}
\label{eq:sigma_cdm}
\sigma^2 \rightarrow \sigma_\mathrm{cdm}^2(z) = \int \frac{\mathrm d k}{k} \frac{k^3 P_\mathrm{cdm}(k,z)}{2 \pi^2} \: W^2(kR) \: ,
\end{equation}
with the cold dark matter power spectrum obtained by rescaling the total matter power spectrum $P_m$ with the respective transfer functions weighted by the density of each species
\begin{equation}
\label{eq:p_cdm}
P_\mathrm{cdm}(k,z) = P_\mathrm{m}(k,z) \left( \frac{\Omega_\mathrm{cdm} T_\mathrm{cdm}(k,z) + \Omega_\mathrm{b} T_\mathrm{b}(k,z)}{T_\mathrm{m}(k,z) (\Omega_\mathrm{cdm} + \Omega_\mathrm{b})} \right)^2 \: ,
\end{equation}
thus assuming that neutrinos are distributed smoothly on cluster scales. The scale dependent growth caused by neutrinos for the other components is also accounted for by the transfer functions. Eq.~\ref{eq:sigma_cdm} is expressed as a time-dependent rescaling, but we can also again think of the inital density field as fixed and map the change to the collapse threshold
\begin{equation}
\label{eq:delta_c_nu}
\delta_c^{\nu} = \frac{\sigma(z)}{\sigma_\mathrm{cdm}(z)} \delta_c \: .
\end{equation}
In this picture, we account for the effect of neutrinos by introducing an appropriate shift in the time variable $\sigma^2$ of the random walk. This rescaling expresses the cold dark matter approximation outlined above and it allows us to compare the effects of modified gravity and neutrinos on the threshold directly. While there is some ambiguity how to compare cosmologies with and without neutrinos, in this paper we choose to keep the total matter density $\Omega_\mathrm{m}$ fixed. Thus when adding neutrinos, we rescale the dark matter density by \citep{Lesgourgues2006}
\begin{equation}
\label{eq:Om_fix_neutrinos}
\Omega_\mathrm{cdm}' = \Omega_\mathrm{cdm} - \frac{\sum m_\nu}{93.14~\mathrm{eV}} \: .
\end{equation}

In Fig.~\ref{fig:delta_c_delta_n_neutrino} we show the rescaled critical density for collapse $\delta_c^\nu$ and the resulting effect on the halo mass function. A larger $\delta_c$ leads to an increased exponential suppression of high mass halos in Eq.~\ref{eq:multiplicity_general_barrier}. Note that the scale dependent growth caused by neutrinos translates to a weak mass dependency of the barrier. To check how this suppression can mask the additional abundance caused by modified gravity, we combine the $f(R)$ threshold with the neutrino rescaling from Eq.~\ref{eq:delta_c_nu}:
\begin{equation}
\label{eq:delta_c_fR_nu}
\delta_c^\mathrm{eff} = \frac{\sigma(z)}{\sigma_\mathrm{CDM}(z)} \delta_c^{f(R)} \: .
\end{equation}
A suitable combination of neutrino masses and $f_{R0}$ can then lead to an effective barrier close to its $\Lambda$CDM value over the mass range $M > 10^{14} M\odot h^{-1}$ relevant for cluster surveys, as demonstrated in Fig.~\ref{fig:delta_c_nu_fR}. We will return to this point and check the validity of this approach by comparing to simulations in Sec.~\ref{sec:calibration}.

\begin{figure}
	\includegraphics[width=\columnwidth]{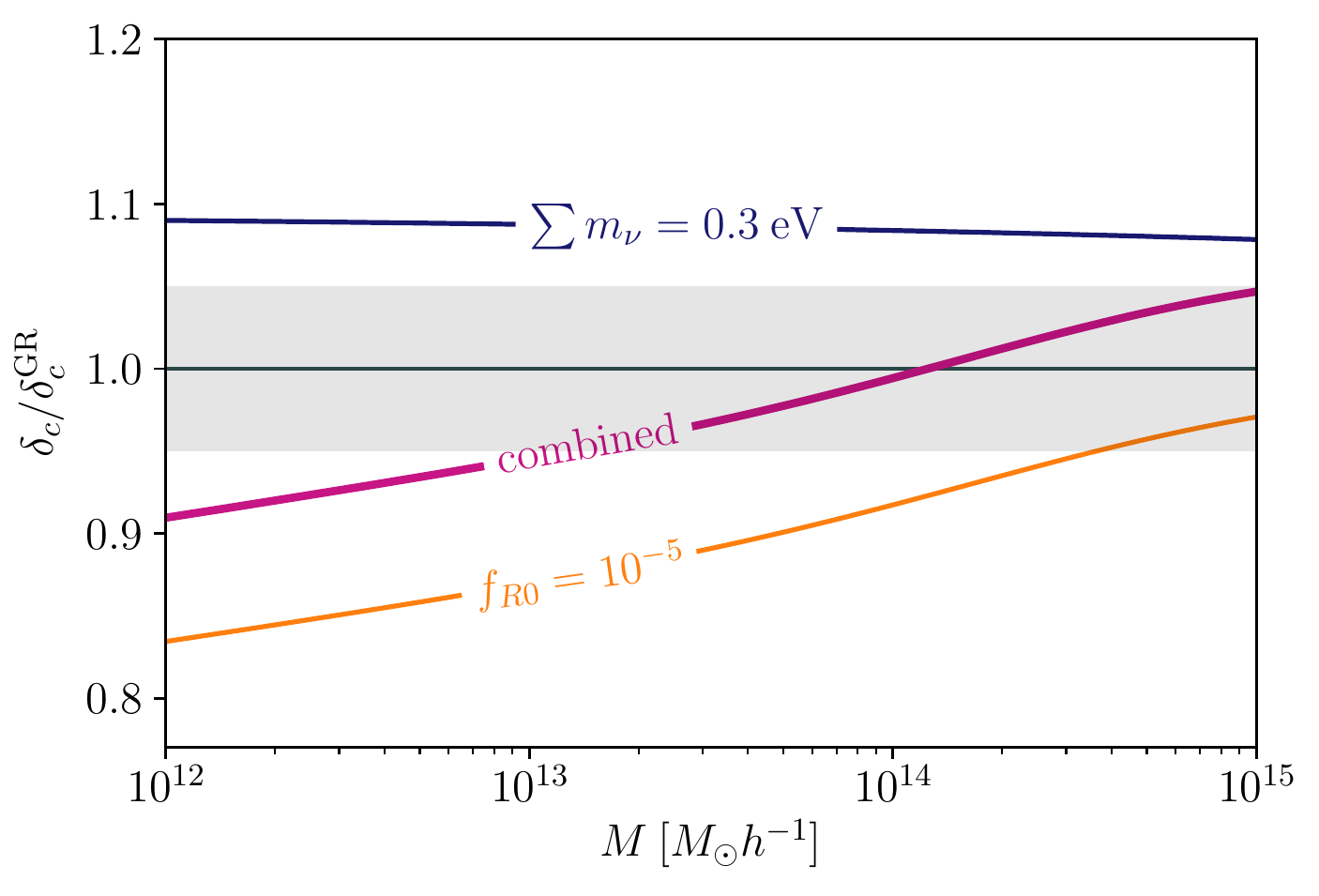}
    \caption{Change in the effective collapse threshold at $z=0$ induced by massive neutrinos with $\sum m_\nu = 0.3 \: \mathrm{eV}$ (blue), for $f_{R0} = 10^{-5}$ (orange) and the combined effect. The grey shaded region shows a $5 \%$ deviation from the fiducial value. Over the mass range $M > 10^{14} M_\odot h^{-1}$ relevant for cluster abundance studies the effects of neutrinos and modified gravity are approximately degenerate.}
    \label{fig:delta_c_nu_fR}
\end{figure}

\subsection{Halo bias and cluster clustering}
\label{sec:clustering_bias}

The mass function also allows us to derive the corresponding clustering bias. The Eulerian bias is given by the overabundance of objects in a region with an overdensity $\delta_0$ compared to the mean abundance
\begin{equation}
b = 1 + \frac{1}{\bar n(M)} \frac{\mathrm d \bar n(M | \delta_0)}{\mathrm d \delta_0} = 1 + \frac{1}{f(\sigma)} \left. \frac{\mathrm d f(\sigma | \delta_0, \sigma_0)}{\mathrm d \delta_0} \right \rvert_{\delta_0 = 0} \: ,
\end{equation}
therefore the first order bias is the linear response of the halo field to changes in the underlying density field. For a fixed barrier, the conditional mass function $f(\sigma | \delta_0)$ simply involves a shift of the barrier $\delta_c \rightarrow \delta_c - \delta_0$, but for a generic barrier the situation is more complicated.

\cite{Achitouv2016} proposed the conditional mass function for a generic barrier
\begin{align}
\label{eq:f_k_conditional}
f(S | \delta_0, S_0) =&~\sqrt{\frac{2}{\pi}} \left( \bar B - S \frac{\mathrm d \bar B}{\mathrm d S} + \frac{S^2}{2} \frac{\mathrm d^2 \bar B}{\mathrm d S^2} - \delta_0 \right) \frac{S/a}{\left( S/a - S_0 \right)^{3/2}} \\
&~ \times \exp \left(-\frac{(\bar B - \delta_0)^2}{2 (S/a - S_0)} \right) \: ,
\end{align}
and found good agreement with Monte Carlo random walks for various barrier shapes. This yields the linear bias
\begin{equation}
\label{eq:linear_bias}
b(S) = 1 + \left( \frac{a \bar B}{S} - \frac{1}{\bar B - S \frac{\mathrm d \bar B}{\mathrm d S}} \right) \: ,
\end{equation}
with the same barrier $\bar B$ as used for the mass function, but the bias depends only mildly on the barrier width $D_B$ and drift $\beta$ for the mass range we focus on in this work. It is mainly sensitive to the mean threshold $\delta_c$.

\begin{figure}
	\includegraphics[width=\columnwidth]{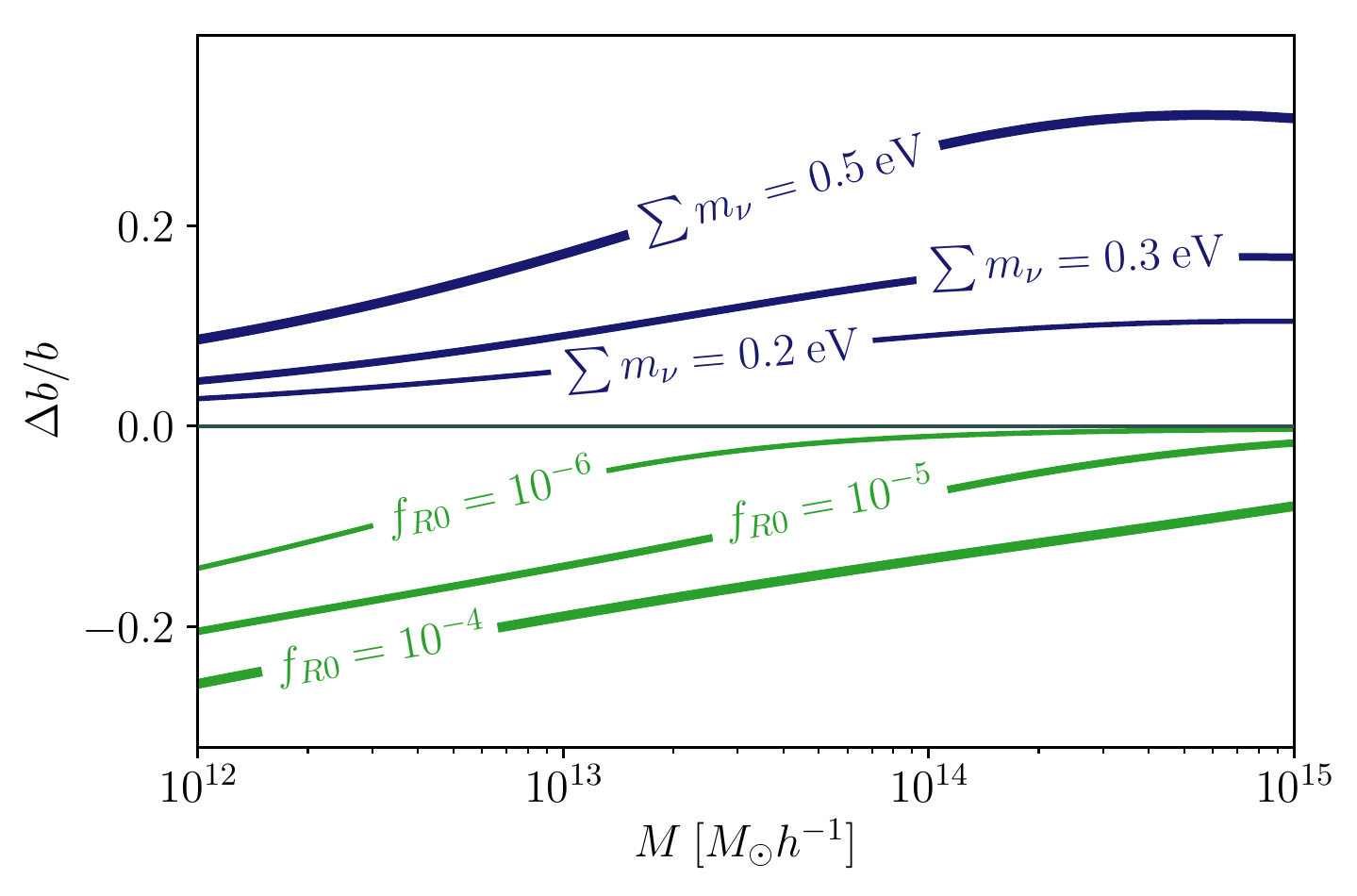}
    \caption{Deviations from the fiducial GR bias for several values of $f_{R0}$ and $\sum m_\nu$ at redshift $z=0$. Clusters become more abundant with larger values of $f_{R0}$, so they become less biased. For neutrinos this trend is reversed, the suppression of high mass objects increases their bias.}
    \label{fig:bias_rel_fR}
\end{figure}

We show the changes in the bias induced by modified gravity or massive neutrinos in Fig.~\ref{fig:bias_rel_fR} using the $f(R)$ barrier $\bar B(M,z,f_{R0})$. The lower threshold means that clusters form out of smaller overdensities compared to $\Lambda$CDM, so they are less biased tracers of the density field. This tendency is only enhanced the stronger the $f(R)$ effect gets and the linear bias shrinks with larger values of $f_{R0}$. For neutrinos, this effect is reversed: because the high-mass tail of the mass function is suppressed, massive clusters are less abundant overall and therefore only form in very overdense regions. 
However the absolute scale of the halo bias in $\Lambda$CDM is still uncertain \citep{Baxter2016, Paech2017}, making it very difficult to use this behaviour for constraints -- both neutrinos and modified gravity lead to a lower bias of low-mass objects compared to high-mass objects.
We therefore leave a forecast analysis also including the clustering of clusters for future work.

\subsection{Calibration and Comparison}
\label{sec:calibration}

The excursion set framework predicts the mass function in terms of the halo mass at virialization $\mathrm d n / \mathrm d M_\mathrm{vir}$ since this is the time at which the halo stops to collapse. 
Moving to modified gravity, the virial overdensity is even more complicated. While constant in an Einstein-de-Sitter universe, we expect $\Delta_\mathrm{vir}^{f(R)}$ to evolve with both redshift and $f_{R0}$.

From the observational point of view, however, the mass of a cluster is often defined as the mass inside a sphere encompassing an overdensity $\Delta$ times a reference value. In this work we adopt $\Delta_m=200$ with respect to the mean matter density as given in Eq.~\ref{eq:M200mean} to define our simulated catalogues and calibrate the mass function accordingly.

\begin{figure}
	\includegraphics[width=\columnwidth]{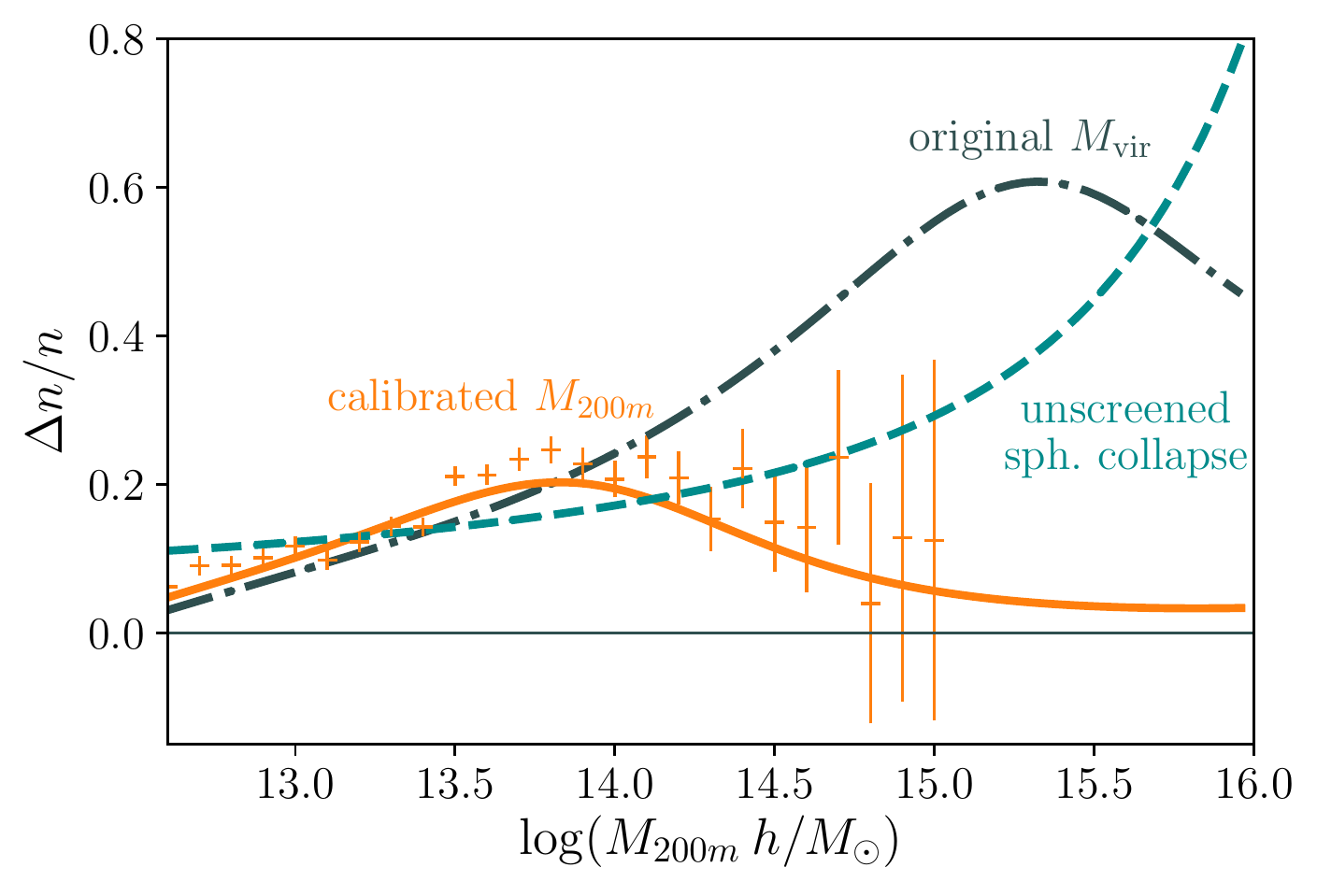}
    \caption{Relative simulated $M_{200m}$ halo abundance for $f_{R0} = 10^{-5}$ at redshift $z=0.3$ compared with the original spherical collapse prediction for virial masses (grey, dot-dashed) and our calibrated model (orange, solid). For comparison, we also show the unscreened spherical collapse prescription from Eq.~\ref{eq:f_x_ST} (cyan, dashed) used in previous studies as explained in the main text.}
    \label{fig:fR_calibration}
\end{figure}

A first exemplary comparison between the fiducial barrier model Eq.~\ref{eq:fR_barrier} and our simulation is shown in Fig.~\ref{fig:fR_calibration} for $f_{R0} = 10^{-5}$ and $z=0.3$. As expected the virial mass function from \cite{Kopp2013} is a bad fit to the $M_{200m}$ catalogue and we can see that the screening mass is offset, leading to a wrong position and amplitude of the $f(R)$ bump. To calibrate the mass function to a new mass definition, we focus on the screening mass $m_b$ in Eq.~\ref{eq:m_b}. We keep the functional form, but because the position and evolution of the screening mass scale is different for another mass definition, we re-fit the parameters $\mu_1, \mu_2$ to account for the evolution with $f_{R0}$, $\beta_3$ to adapt the height of the relative abundance peak and $\alpha_4$ to adjust the redshift evolution. This is done via minimisation of the Gaussian log-likelihood

\begin{align}
\label{eq:log-lkl_calibration}
\ln \mathcal L =& ~-\frac{1}{2} \left(\mathbf N^\mathrm{theo} - \mathbf N^\mathrm{sim} \right) \mathbf C^{-1} \left(\mathbf N^\mathrm{theo} - \mathbf N^\mathrm{sim} \right)^T \\
& ~- \frac{1}{2} \ln \det  \mathbf C^{-1} \nonumber \: ,
\end{align}
where the covariance matrix consists of a Poissonian contribution and a sample variance term
\begin{equation}
C^{-1}_{ij} = \delta_{ij} N_i^\mathrm{theo} + b_i b_j N_i^\mathrm{theo} N_j^\mathrm{theo} \sigma(V_\mathrm{box}) \:
\end{equation}
with theoretical cluster counts $N_i^\mathrm{theo}$ per mass bin $i$ and $\sigma(V_\mathrm{box})$ is the variance of the density field computed inside the box. We calculate the mean bias averaged over a bin $\Delta M_i$ as
\begin{equation}
\bar b_i = \int_{\Delta M_i} \mathrm d M \frac{\mathrm d n}{\mathrm d M} b(M) \Big/ \int_{\Delta M_i} \mathrm d M \frac{\mathrm d n}{\mathrm d M} \: ,
\end{equation}
using Eq.~\ref{eq:linear_bias} for the bias and Eq.~\ref{eq:f_x_fR} for the mass function. Note that the barrier shape given by $D_B$ and $\beta$ is very important for the proper GR limit, but largely cancels in Eq.~\ref{eq:f_x_fR}. The mass function ratio is therefore almost completely independent from the fiducial barrier values. So while we choose to work within a consistent framework with a mass function that is extended to $f(R)$, one could also replace $f_x^\mathrm{GR}$ in Eq.~\ref{eq:f_x_fR} with another multiplicity function such as ones by \cite{Tinker2008} or \cite{Crocce2010} as long as it is also calibrated to $M_{200m}$. We do not perform a comprehensive comparison of mass functions here, but we note that our results for bias and multiplicity agree within $\sim 5 \%$ with those established results in the literature -- a value we take as an estimate for current systematic effects on the halo mass function mainly due to differences in halo definition.

\begin{figure*}
	\includegraphics[width=2\columnwidth]{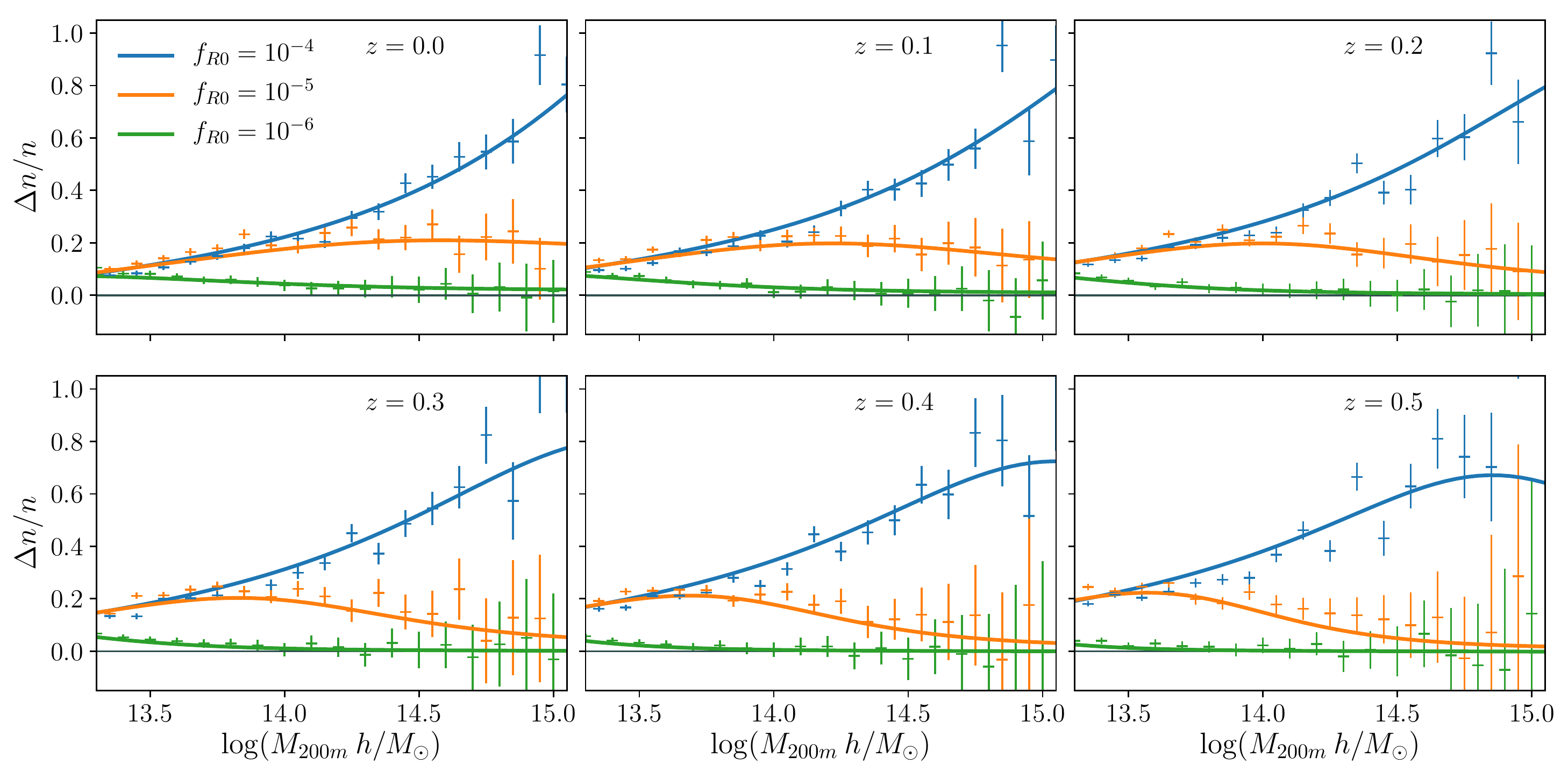}
    \caption{Calibrated halo mass function ratio for various redshifts and $f_{R0} = 10^{-4}$ (blue), $10^{-5}$ (orange) and $10^{-6}$ (green) compared to our simulation suite. The $f(R)$ bump in the relative abundance moves towards lower masses with redshift.}
    \label{fig:fR_ratio}
\end{figure*}

Within our simulations, we find no preference for any redshift evolution in the GR barrier parameters $D_B$ and $\beta$. We fit them to our $\Lambda$CDM simulations first and keep them fixed while calibrating the remaining $f(R)$ parameters $\alpha_4,~\beta_3,~\mu_1$ and $\mu_2$ to our fR4, fR5 and fR6 simulations. The resulting best-fit values with statistical errors are shown in Tab.~\ref{tab:fit_results}. For the $\Lambda$CDM barrier values we find qualitative agreement with previous similar studies \citep{MaggioreII,Kopp2013,Achitouv2016} while the position and evolution of the screening mass $m_b$ given by the other parameters deviates substantially from the virial mass function from \cite{Kopp2013}. The results are compared to our simulated catalogues in Fig.~\ref{fig:fR_ratio} for a wide range of redshifts and values of $f_{R0}$.
We find that our model for the halo mass function can reproduce the simulated data by fitting only four parameters to account for the full non-linear behaviour of the modified gravity model.

\begin{table}
	\centering
	\caption{Best-fit parameters for the width $D_B$ and the drift $\beta$ of the fiducial GR barrier and the calibrated values for the modified gravity threshold $\delta_c(f_{R0}, z)$ in Eq.~\ref{eq:fR_barrier}. The statistical uncertainty of the fit is the last significant digit.}
	\label{tab:fit_results}
	\begin{tabular}{c c | c c c c} % six columns, alignment for each
		GR & & $f(R)$ & & & \\
        \hline \hline
		$D_B$ & $\beta$ & $\alpha_4$ & $\beta_3$ & $\mu_1$ & $\mu_2$\\
		$0.37 $  & $0.11 $ & $0.067 $ & $5.6 \times 10^{-3}$ & $1.38 $ & $21.32$ \\
        \hline
	\end{tabular}
\end{table}

For completeness, we also compare our result to a previously used prescription for the modified gravity mass function in Fig.~\ref{fig:fR_calibration}. In this ansatz proposed by \cite{Cataneo2014}, the relative effect of $f(R)$ is captured by a ratio of ellipsoidal collapse multiplicity functions
\begin{equation}
\label{eq:f_x_ST}
\frac{f^{f(R)}}{f^\mathrm{GR}} \approx \frac{f^{\mathrm{ST}}(\sigma^{f(R)}, \delta_{c, \mathrm{unscr.}}^{f(R)})}{f^{\mathrm{ST}}(\sigma^{\mathrm{GR}}, \delta_c^{\mathrm{GR}})} \: ,
\end{equation}
where $f^{\mathrm{ST}}$ denotes the mass function by \cite{Sheth2002}. The density variance is calculated using the linear power spectrum $P(k)$ in the respective theory, and $\delta_{c, \mathrm{unscr.}}^{f(R)}$ denotes the threshold for spherical collapse in case the theory is unscreened everywhere, i.e. gravity is enhanced by $4/3$ \citep{Schmidt2009}
\begin{equation}
\label{eq:delta_c_fR_unscreened}
\delta_{c,\mathrm{unscr.}}^{f(R)}(z) = 1.7063 \left(1 - 0.0136 \log_{10} \bigg( 1 + \frac{\Omega_m ^ {-1} - 1}{(1 + z)^3} \bigg) \right) \: ,
\end{equation}
which shares the functional form of Eq.~\ref{eq:delta_c_GR} but differs in the numerical coefficients. In comparison to our simulations, we can see that this prescription fails to properly predict the onset and shape of the characteristic enhancement peak.

The next step is to test the inclusion of neutrinos into our framework via Eq.~\ref{eq:delta_c_fR_nu}. We show the combined effect of neutrinos and modified gravity measured from our simulations in Fig.~\ref{fig:fR_nu_ratio} - note that the simulations including neutrinos were not used to fit the mass function parameters. Both cosmologies show an approximate degeneracy leading to an abundance of clusters that is within $10 \%$ consistent with $\Lambda$CDM expectation at $z=0$, and the behaviour is well captured by our mass function. This cancellation weakly depends on redshift, so cosmologies with similar mass functions at $z=0$ will in general differ at earlier times. The precise degeneracy depends on the survey specifications such as redshift range and selection function, and we will return to this problem within the full cosmological parameter space in the next section.

\begin{figure}
	\includegraphics[width=\columnwidth]{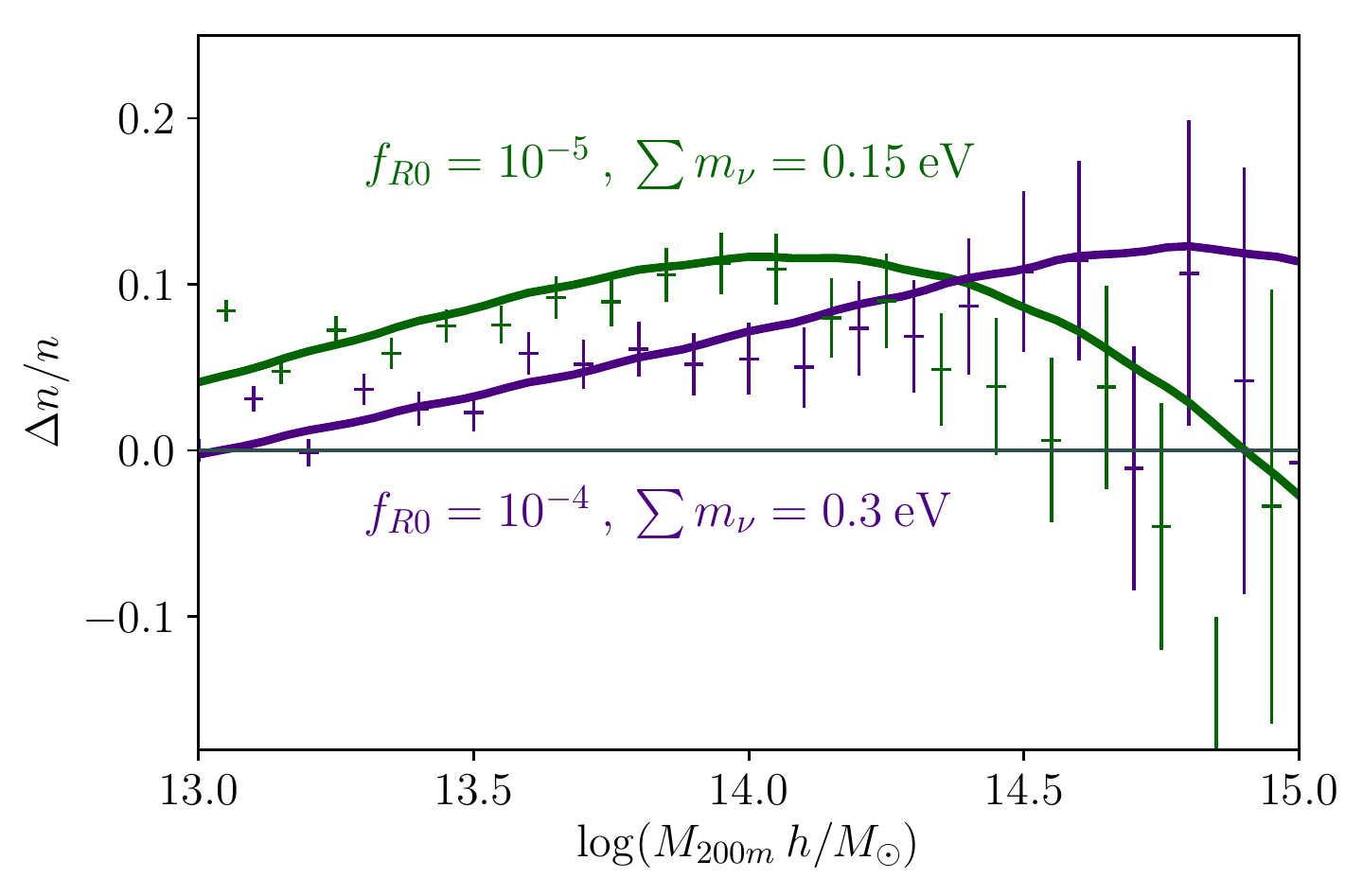}
    \caption{Joint effect of modified gravity and neutrinos on the relative halo abundance at $z=0$. The theoretical abundance is calculated by combining the calibrated $f(R)$ barrier with the neutrino rescaling (Eq.~\ref{eq:delta_c_nu}). Both cluster abundance predictions deviate by less then $10 \%$ from the $\Lambda$CDM predictions.}
    \label{fig:fR_nu_ratio}
\end{figure}

\section{Forecasts}
\label{sec:forecasts}

To assess if differences in the cluster abundance are measurable, it is important to consider the changes in the halo mass function in the context of a survey with a specific selection function.

We now show with two idealised test cases the consequences of our results for the ability of current and future surveys to constrain $f_{R0}$. The abundance of clusters is mostly sensitive to $(\Omega_m, \sigma_8, \sum m_\nu, \log f_{R0})$;
as for the other relevant cosmological parameters we include priors from different probes.
This has to be done with caution, because datasets might show different results when analysed in a $f(R)$ framework. We therefore make use of the fact that the model reproduces a $\Lambda$CDM expansion history and limit ourselves to \textit{geometrical} probes.

We add baryon acoustic oscillation priors on the distance scale $D_V(z)$ based on BOSS DR12 data \citep{BOSS_DR12_2017} at redshifts $z = 0.38,~0.51$ and $0.61$. We centre them on our fiducial cosmology and assume pre-reconstruction errors on the data points, i.e. without assuming a $\Lambda$CDM model to linearise the BAO signal, which results in conservative results. We denote this data set with BAO. Complementary, big-bang nucleosynthesis measurements constrain the baryon density $\Omega_b h^2$ in the early universe, where any $f(R)$ effects are negligible. The width of the error bar is based on \cite{BBN_Cooke2014}. A summary of both sets of Gaussian priors is given in Tab.~\ref{tab:priors}.

\begin{table}
	\centering
	\caption{A summary of the complementary BAO and BBN mock data sets used in combination with cluster counts if indicated. We assume these result in Gaussian priors on the measured quantity with mean $\mu$ and width $\sigma$.}
	\label{tab:priors}
	\begin{tabular}{l c | c c} 
		Probe & Quantity & $\mu$ &$\sigma$ \\
        \hline \hline
		BAO & $D_V(z=0.38) / r_s$ & 10.05 & $0.17$ \\
 		 & $D_V(z=0.51) / r_s$ & 12.84 & $0.13$ \\
		 & $D_V(z=0.61) / r_s$ & 14.77 & $0.13$ \\
       \hline
		BBN & $100 \times \Omega_b h^2$ & 2.224 & $0.046 $\\
        \hline
\end{tabular}
\end{table}

The most powerful complementary data set comes from the CMB. If indicated, we combine the cluster data with priors on the primary CMB parameters derived from the Planck-high-$\ell$ temperature power spectrum. We use the publicly available chains either for the base model or including varying neutrino masses to derive the covariance matrix and use this Gaussian prior, again centred on our fiducial cosmology. While changes to the temperature anisotropy power spectrum by $f(R)$ gravity are introduced via the integrated Sachs-Wolfe effect at late times, the impact on multipoles $\ell > 30$ is very small for the relevant parameter space.

\subsection{Optical cluster surveys}
\label{sec:forecast_DES}

We now explore these effects in the context of a forecast for a optical cluster survey, where the main observable is the cluster richness $\lambda$. We model the expected number counts per bin in redshift $\Delta z_i$ and richness $\Delta \lambda_j$ as
\begin{equation}
\langle N_{ij} \rangle = \Omega \int_{\Delta z_i} \mathrm d z \frac{\mathrm d V}{\mathrm d z} \int_0^\infty \mathrm d M \frac{\mathrm d n}{\mathrm d M} \int_{\Delta \lambda_j} \mathrm d \lambda \: p(\lambda | M) \: ,
\end{equation}
where the survey area $\Omega$ is fixed, and introduce the probability $p(\lambda | M)$ for a cluster of mass $M$ to be observed with a richness $\lambda$. We assume a log-normal distribution, which allows us to solve the integration over the observable to arrive at
\begin{equation}
\langle N_{ij} \rangle = \Omega \int_{\Delta z_i} \mathrm d z \frac{\mathrm d V}{\mathrm d z} \int_0^\infty \mathrm d M \frac{1}{2} \bigg(\erfc(x_\mathrm{min}) - \erfc(x_\mathrm{max})  \bigg) \frac{\mathrm d n}{\mathrm d M} \: ,
\end{equation}
with
\begin{equation}
x_\mathrm{min/max} \equiv \frac{\ln \lambda_\mathrm{min/max} - \langle \ln \lambda \rangle(M)}{\sqrt[]{2 \sigma^2_{\ln \lambda} }} \: .
\end{equation}
We use the weak-lensing calibrated $M - \lambda$ relation measured by \cite{Murata2018} on SDSS clusters:
\begin{align}
\langle \ln \lambda \rangle (M) &= A + B \ln \left( \frac{M}{M^\star} \right) \\
\sigma_{\ln \lambda} (M) &= \sigma_0 + q \ln \left( \frac{M}{M^\star} \right) \: ,
\end{align}
where $M^\star = 3 \times 10^{14} M_\odot / h$ is the pivot mass of the relation and $A, B, \sigma_0$ and $q$ are free parameters varied within priors given by the measurements by \cite{Murata2018}. Note that the weak lensing mass estimate of a given cluster is not affected by $f(R)$ because geodesics are unchanged up to a negligible factor $1 + f_{R0}$.

In addition to these observational uncertainties, also the mass function measured in simulations shows systematic scatter. This is mainly caused by ambiguities in the halo definition, so even an identical underlying dark matter field can result in slightly different halo statistics. Typically, different halo finders vary in the resulting amplitude and tilt of the mass function \citep{Knebe2011}, so we assume
\begin{equation}
\frac{\mathrm d n}{\mathrm d M} \rightarrow \frac{\mathrm d n}{\mathrm d M} \left(\gamma + \eta \log \bigg( \frac{M}{M^\star} \bigg) \right)
\end{equation}
with $\gamma$ and $\eta$ free to vary with Gaussian priors with width $\sigma = 0.05$ centred at 1 and 0 respectively. Because these systematic errors are by far larger than statistical uncertainty in our fit of barrier parameters, we keep the latter fixed.

The selection function is crucial for the specific degeneracy between parameters, so we distinguish two cases: Either a large, shallow layout or a deeper survey focused on a smaller sky area.

\begin{figure}
	\centering
	\includegraphics[width=0.9\columnwidth]{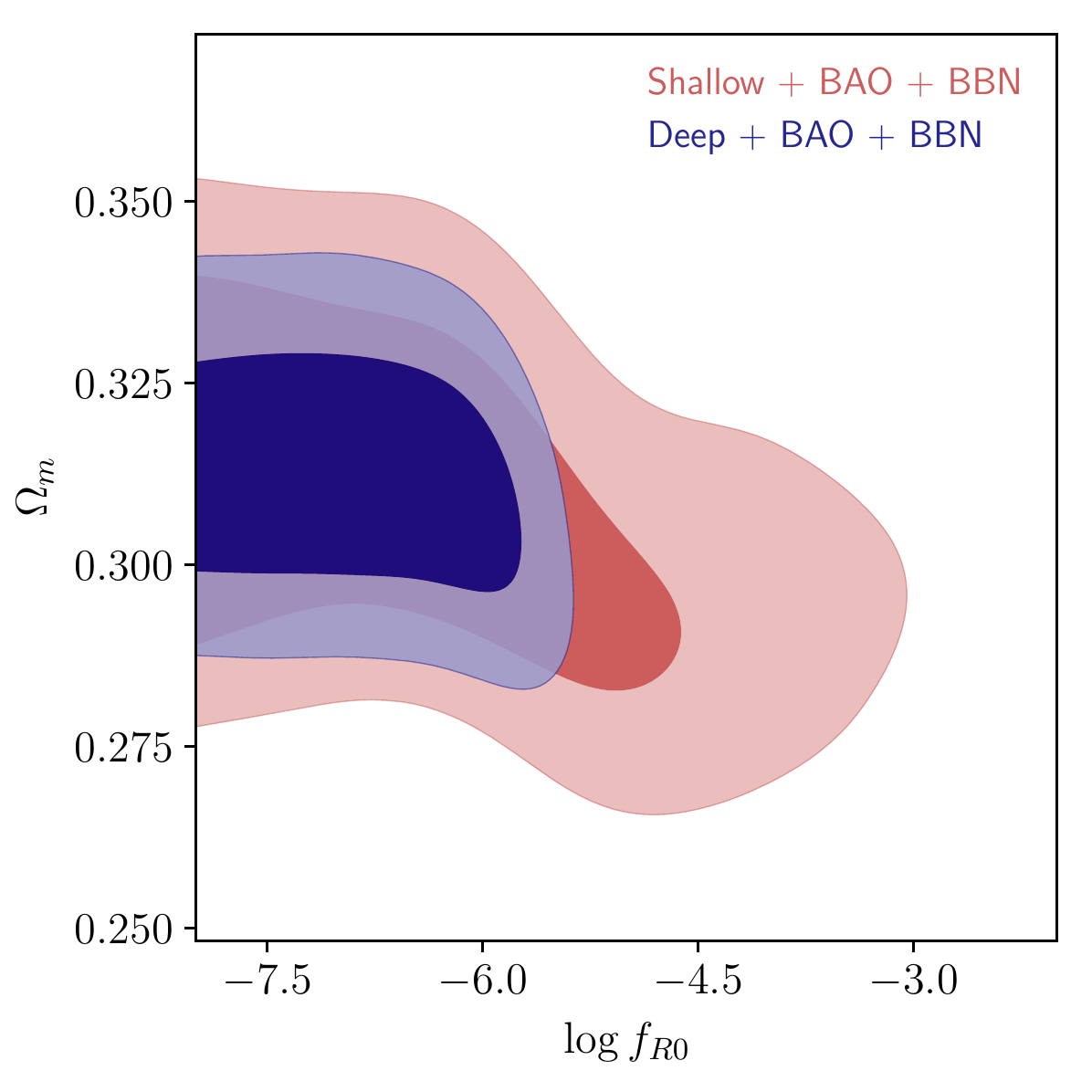}
    \caption{Expected constraints on $f_{R0}$ from a shallow or deep optical cluster survey as described in the text. The cases shown here keep the neutrino mass fixed.}
    \label{fig:optical_2d_forecast}
\end{figure}

For the shallow case, we assume an area of $10^{4}~\mathrm{deg}^2$ with eight richness bins as in \cite{Murata2018} $\lambda \in [20,~25,~30,~35,~40,~47.5,~55,~77.5,~100]$ and one redshift bin $z \in [0.1,~0.3]$. This translates to an approximately flat limiting mass of $M_{\rm min} \sim 10^{14.4} M_\odot/h$. All bins are well populated with over 100 clusters so we assume a Gaussian likelihood as in Eq.~\ref{eq:log-lkl_calibration}. This mock survey is combined with either CMB or BAO + BBN priors as given in Tab.~\ref{tab:priors} and we evaluate the resulting likelihood using the Monte Carlo Markov Chain code \texttt{MontePython} \citep{MontepythonI,MontepythonII}.

\begin{figure*}
	\includegraphics[width=2\columnwidth]{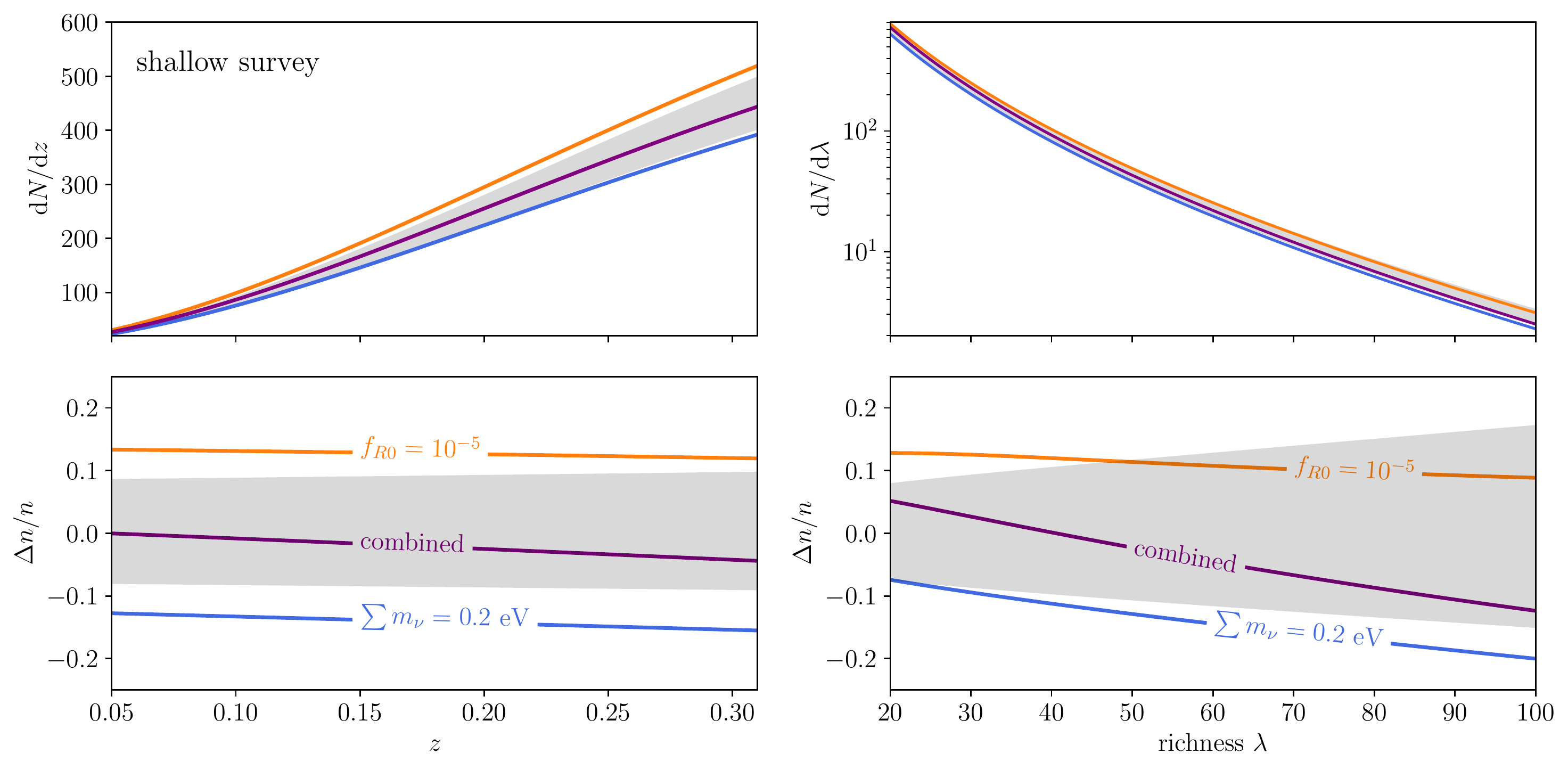}
    \caption{Left: Evolution of cluster counts with redshift for the shallow optical cluster survey described in the text. Grey shaded bands indicate $1\sigma$ uncertainty in the mass-richness-relation. The bottom plot shows relative deviations caused by modified gravity (orange), massive neutrinos (blue), or both (violet). For low redshifts with the given selection function, both effects are approximately a shift in total amplitude of the counts. Right: Richness distribution of cluster counts. The bottom plot shows relative deviations.}
    \label{fig:dndz_dndlambda_shallow}
\end{figure*}

\begin{figure*}
	\includegraphics[width=2\columnwidth]{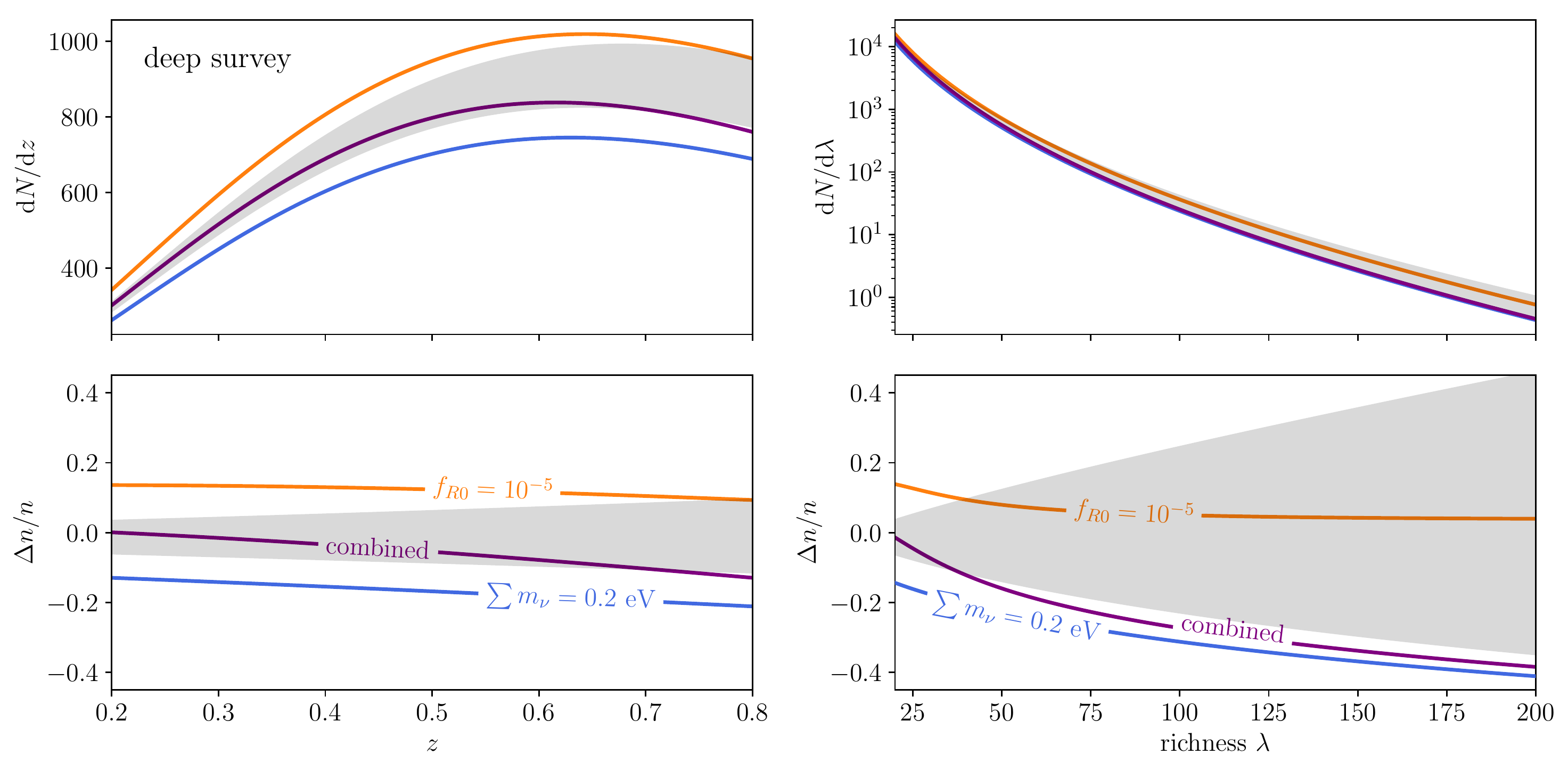}
    \caption{Left: Evolution of cluster counts with redshift for the deep optical cluster survey described in the text. Grey shaded bands indicate $1\sigma$ uncertainty in the mass-richness-relation. The bottom plot shows relative deviations caused by modified gravity (orange), massive neutrinos (blue), or both (violet). While degenerate at low redshifts, neutrino effects are more pronounced at high $z$ (where the $f(R)$ mass function reverts to GR). Right: Richness distribution of cluster counts. The bottom plot shows relative deviations. The degeneracy here crucially depends on the position of the $f(R)$ peak in the relative abundance.}
    \label{fig:dndz_dndlambda_deep}
\end{figure*}

We show the cluster count distribution in redshift and richness for a shallow survey in Fig.~\ref{fig:dndz_dndlambda_shallow}. For the given selection function, at low redshifts the effects of neutrinos and modified gravity are almost completely degenerate. Both roughly translate into a shift in the overall amplitude which is also easily mimicked by the amplitude of the $M - \lambda$ relation. The richness information does help to break this degeneracy slightly because neutrinos tend to cause a strong suppression of very massive clusters while modified gravity leads to a higher abundance of low- and intermediate mass objects. The resulting limits on $f_{R0}$ that can be achieved with such a survey are shown in Tab.~\ref{tab:constraints}. If cluster counts are only combined with BAO information, the limits are rather weak and when adding neutrinos we find no relevant upper bound. Adding the CMB improves the situation by pinning down the other cosmological parameters, but even then adding neutrinos weakens the bounds considerably. Note that there is a small additional effect due to broader CMB constraints on other parameters in a $\nu$CDM cosmology, but this mostly extends the contours in the direction of larger allowed $\Omega_m$ values while $f_{R0}$ is anti-correlated with the matter density.

For the deep survey, we take an area of $5000~\mathrm{deg}^2$ -- the total area that will be covered by the Dark Energy Survey\footnote{https://www.darkenergysurvey.org} -- and bins in richness $\lambda \in [20,~30,~45,~60,~200]$ and redshift $z \in [0.2,~0.35,~0.5,~0.65,~0.8]$. The resulting cluster counts for this configuration in redshift and richness are shown in Fig.~\ref{fig:dndz_dndlambda_deep}. Information about the abundance at higher redshifts helps in breaking the degeneracy, because while neutrinos suppress the population there, the $f(R)$ mass function reverts to GR for $z > 0.5$. Even though modified gravity boosts the abundance of high mass clusters at low redshifts as shown in Fig.~\ref{fig:fR_ratio}, integrated over $z$ the effect on low-richness clusters is dominant as shown on the right panel of Fig.~\ref{fig:dndz_dndlambda_deep}. Neutrinos on the other hand suppress the high-mass end of the halo mass function, so that -- when combined -- the two effects largely break the degeneracy between $f(R)$ and neutrinos. Even without adding CMB information, such a survey can constrain $f_{R0}$ down to the effective cluster floor of $\sim 10^{-6}$ independent of neutrinos. We show the resulting posterior from both surveys combined with BAO and BBN priors for vanishing neutrino mass in Fig.~\ref{fig:optical_2d_forecast}.

\begin{table}
	\centering
	\caption{Forecasted constraints from optical cluster surveys in various configurations as described in the text.}
	\label{tab:constraints}
	\begin{tabular}{l c} 
		Probes & Limit (95 $\%$)\\
        \hline \hline
        Shallow + BAO + BBN & $|f_{R0}| < 8.1 \times 10^{-4}$ \\
        Shallow + BAO + BBN + $\nu$ &  -- \\
		Shallow + BAO + BBN + CMB & $|f_{R0}| < 7.6 \times 10^{-5}$ \\
 		Shallow + BAO + BBN + CMB + $\nu$ & $|f_{R0}| < 1.5 \times 10^{-4}$ \\
		Deep + BAO + BBN + $\nu$ & $|f_{R0}| < 2.0 \times 10^{-6}$  \\
        \hline
\end{tabular}
\end{table}

\subsection{SZ Cluster surveys}
\label{sec:forecast_SZ}

The thermal Sunyaev-Zeldovitch (SZ) effect is the heating of CMB photons by scattering with hot electron plasma in clusters of galaxies, leading to a characteristic distortion of the blackbody spectrum. The measured amplitude is expressed by the Compton $y$-parameter and is given by the integrated electron density $n_e$ weighted with their temperature $T_e$ along the line of sight
\begin{equation}
y \propto \int n_e T_e \mathrm d l \propto M \langle T_e \rangle \: .
\end{equation}
If we assume a virialised system, $\langle T_e \rangle \propto M^{2/3}$ and the amplitude scales as $y \propto M^{5/3}$. The potential energy of such a cluster is given by
\begin{equation}
\label{eq:Epot_Comptony}
\langle E_\mathrm{pot }\rangle \propto - \frac{G M^2}{R} \propto - G M ^{5/3} \propto - y \: ,
\end{equation}
therefore the thermal SZ effect is a probe of the potential energy. In unscreened $f(R)$ gravity, potentials are deeper by a factor of $4/3$ and thus a cluster with the same mass will induce a larger SZ signal compared to a standard cosmology.

A SZ selected cluster sample will hence show a higher abundance in modified gravity both due to the mass function enhancement discussed so far, but also due to modifications of the selection function because lower mass clusters will surpass the detection threshold.

\begin{figure}
	\includegraphics[width=\columnwidth]{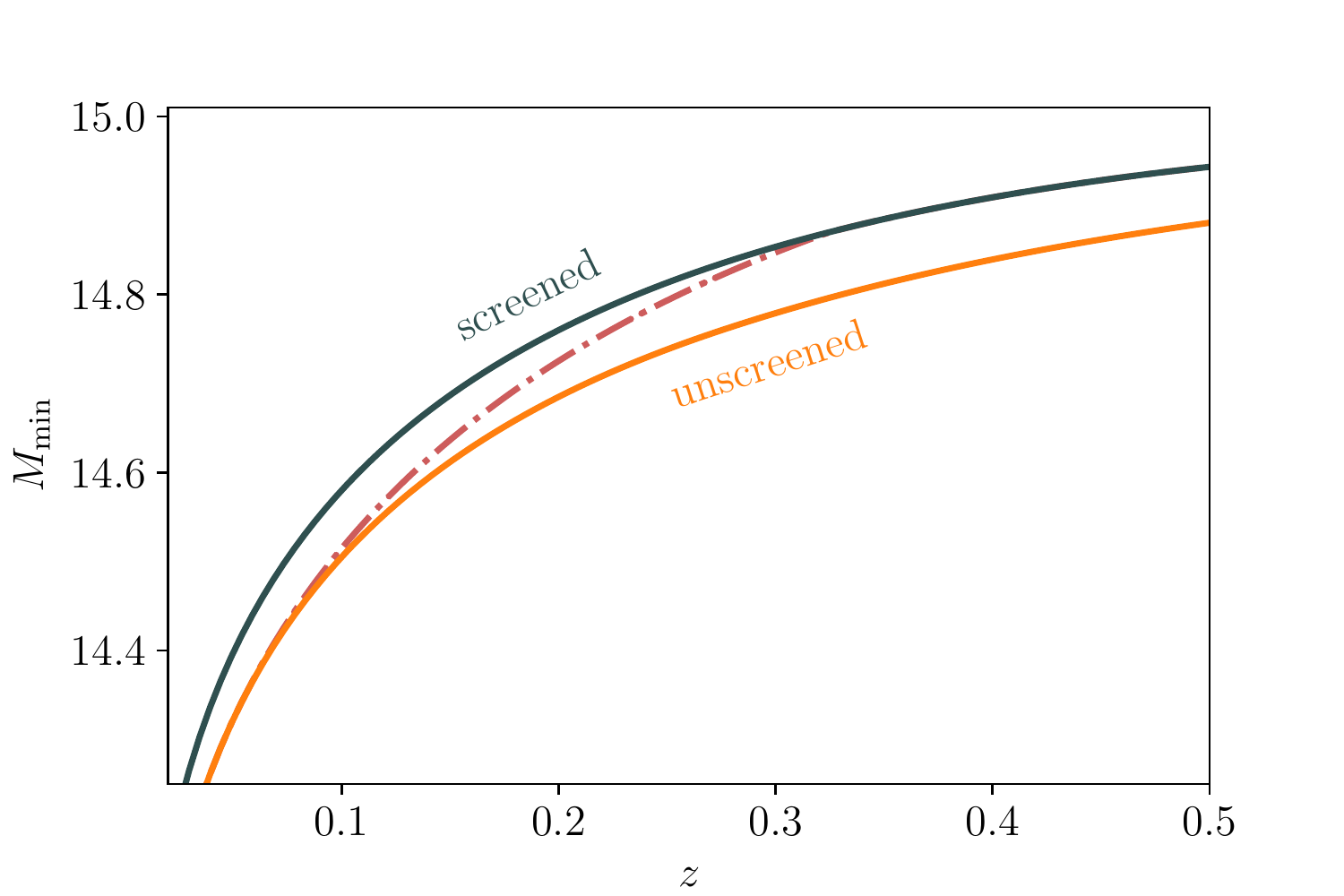}
    \caption{Limiting mass using the Planck SZ cluster selection function for the fiducial case (grey), assuming all clusters are unscreened for a high value of $f_{R0} = 10^{-4}$ (orange) and an intermediate case where parts of the sample are screened with $f_{R0} = 5 \times 10^{-5}$ (red, dot-dashed).}
    \label{fig:planck_selection}
\end{figure}

To model this effect, we consider the relative strength of gravity
\begin{equation}
g(r) \equiv \frac{\mathrm d \psi / \mathrm d r}{\mathrm d \psi_N / \mathrm d r}
\end{equation}
normalised by the Newtonian expectation $\psi_N$ which varies between 1 in the screened regime and 4/3 for the unscreened case. From this we can derive the weighted average
\begin{equation}
\label{eq:g_bar}
\bar g = \frac{\int \mathrm d r r^2 w(r) g(r)}{\int \mathrm d r r^2 w(r)} \: ,
\end{equation}
with the weighting function
\begin{equation}
w(r) = \rho(r) r \frac{\mathrm d \psi_N}{\mathrm d r}
\end{equation}
which corresponds to the averaged additional potential energy. We follow \cite{Schmidt_2010} and make the simplified assumption that the fifth force is only sourced by mass outside of the radius given by Eq.~\ref{eq:thin_shell}. Therefore we write
\begin{equation}
g(r) = 1 + \frac{1}{3} \frac{M(<r) - M(<r_\mathrm{screen})}{M(<r)} \: ,
\end{equation}
where $r_\mathrm{screen}$ is the radius where the equality in Eq.~\ref{eq:thin_shell} holds.  The time evolution of $r_\mathrm{screen}$ and subsequently $\bar g$ is induced by the background evolution of $f_R$
\begin{equation}
\bar f_R(z) = f_{R0} \frac{1 + 4 \frac{\Omega_\Lambda}{\Omega_m}}{(1+z)^3 + 4 \frac{\Omega_\Lambda}{\Omega_m}} \: ,
\end{equation}
and the integrals in Eq.~\ref{eq:g_bar} can be solved by assuming NFW profiles so both density and potential are determined. Note that $\bar g$ is only very weakly sensitive to the concentration of the profiles, so we fix the relation to the results of \cite{Bullock2001}. Even though halos tend to be more concentrated in $f(R)$, this does not change our qualitative argument.

From Eq.~\ref{eq:Epot_Comptony} we therefore expect the mass estimate to be biased compared to GR by
% Because the potential energy then scales as $\propto M^{5/3}$, we expect the effective mass estimate from SZ measurements to be biased by \citep{Schmidt_2010}
\begin{equation}
M_\mathrm{eff} = \bar g^{3/5} M_\mathrm{true} \: ,
\end{equation}
i.e. the SZ signal coming from an unscreened cluster of fixed mass is higher by a factor of $(4/3)^{3/5} \simeq 1.19$ compared to the GR expectation. Similar arguments have been used before to constrain $f(R)$ by comparing lensing masses with X-ray \citep{Wilcox2015} or dynamical mass estimates \citep{Pizzuti2017}. Here we want to incorporate the effect into a cluster abundance framework.

\begin{figure}
	\includegraphics[width=\columnwidth]{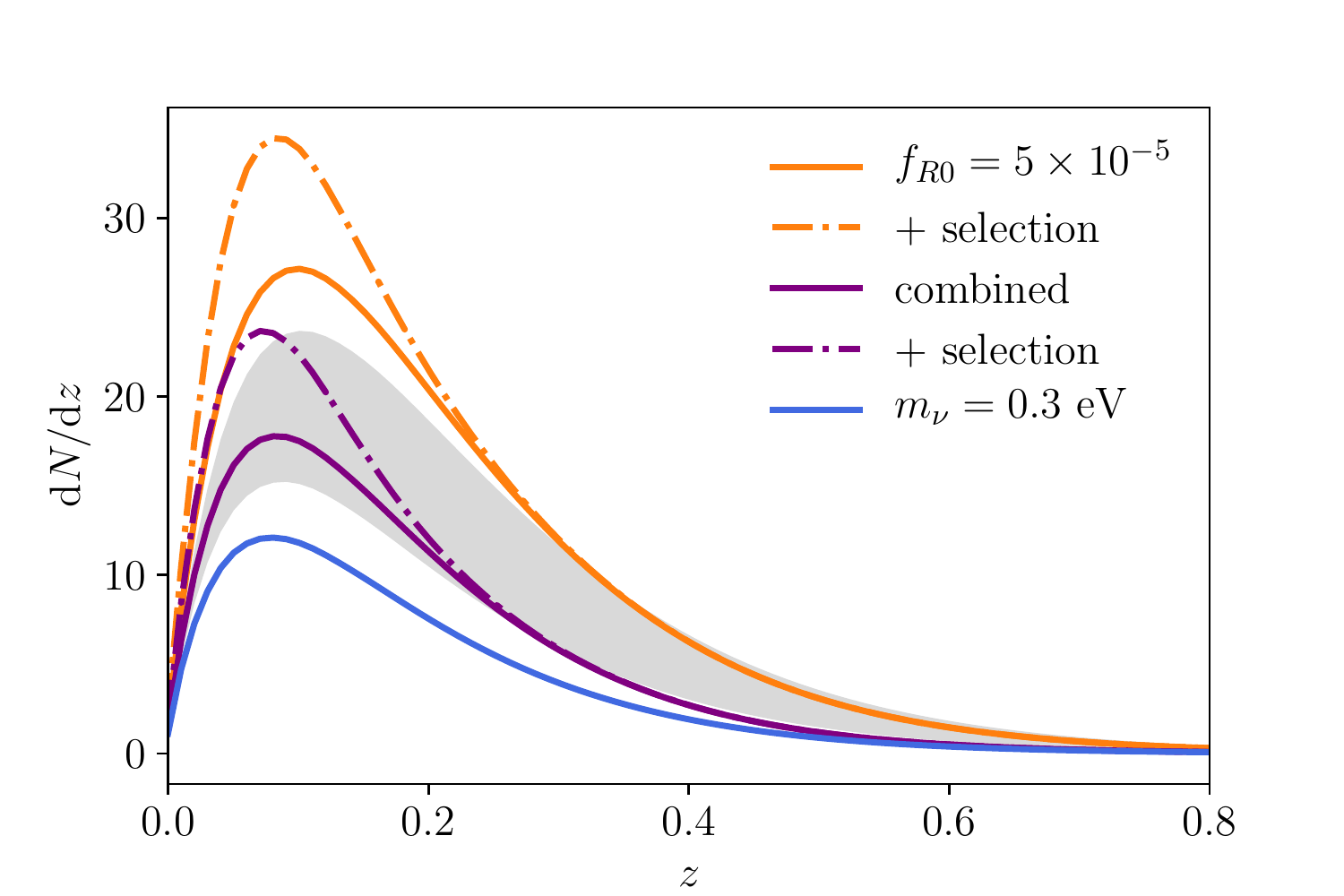}
    \caption{Redshift distribution for the Planck cluster counts due the halo mass function into account (solid) and selection function effects (dot-dashed). Grey bands indicate a $10\%$ uncertainty in the cluster mass scale $(1-b)$.}
    \label{fig:planck_dndz}
\end{figure}

To illustrate the method, we consider the consequences for the Planck SZ cluster sample \citep{PlanckXXIV}. There, the hydrostatic mass bias $(1-b)$ is introduced to account for the difference between masses inferred from lensing and the corresponding hydrostatic SZ signal. In $f(R)$, we therefore expect $(1-b)$ to be modified by an additional factor $\bar g^{3/5}$. Because the mass definition used in SZ surveys is typically $M_{500c}$, we calculate NFW potentials to determine $\bar g$ using this mass definition and we consider a cluster fully screened if the condition in Eq.~\ref{eq:thin_shell} has not been met at $R_{500c}$.

In Fig.~\ref{fig:planck_selection} we show the resulting limiting mass for the Planck SZ selection function. Because the clusters in the sample are very massive, they are screened unless $f_{R0}$ reaches quite high values $\sim 10^{-4}$. However, if all clusters in the Planck sample are unscreened, this would be completely absorbed by the fiducial measurement of the bias factor - but because the lensing calibration is performed on very massive objects, smaller objects can still exhibit deviations. This is illustrated with the dot-dashed line for $|f_{R0}| = 5 \times 10^{-5}$.

The resulting Planck SZ cluster counts are shown in Fig.~\ref{fig:planck_dndz}. Here we recalibrate our mass function to $M_{500c}$ using the rescaling outlined in \cite{Hu2003}. While this simplified procedure will not predict the position of the screening mass and the subsequent position of the $f(R)$ peak in the mass function correctly, we just want to point out that the effect of the adjusted selection function can be quite powerful - in this case as important as the higher cluster abundance from the mass function itself.

The high mass scale for the Planck clusters limits the usefulness of this method here, but upcoming X-ray surveys such as \textit{eRosita}\footnote{http://www.mpe.mpg.de/eROSITA} are expected to detect clusters and groups down to $M \sim 10^{13} M_\odot / h$ where similar methods can be very powerful.

\subsection{Searching for modified gravity with other parametrisations}

The problem in searching for modifications of gravity is that theory space is enormous, and there are potentially many models to test. Current and future cosmological surveys are mostly designed to search for deviations in the dark energy equation of state $w$ from $-1$, so we might wonder if these standard searches are sufficient to detect deviations from $\Lambda$CDM without assuming a specific model. The hope is then that once an anomaly is detected (for example an equation of state $w \neq -1)$, one can resolve the tension in an extended model involving new physics.

As a test case, we set $\sum m_\nu = 0$ and generate a fiducial cluster catalogue with $f_{R0} = 10^{-4}$ for the shallow optical cluster survey described above combined with CMB and BAO + BBN information. This value of $f_{R0}$ is larger than the 95 \% upper limit $f_{R0} < 7.2 \times 10^{-5}$ from the same combination of data sets given in Tab.~\ref{tab:constraints}. We then explore the posterior assuming a $w$CDM model and use the according CMB covariance matrix for our prior.

\begin{figure}
	\includegraphics[width=1\columnwidth]{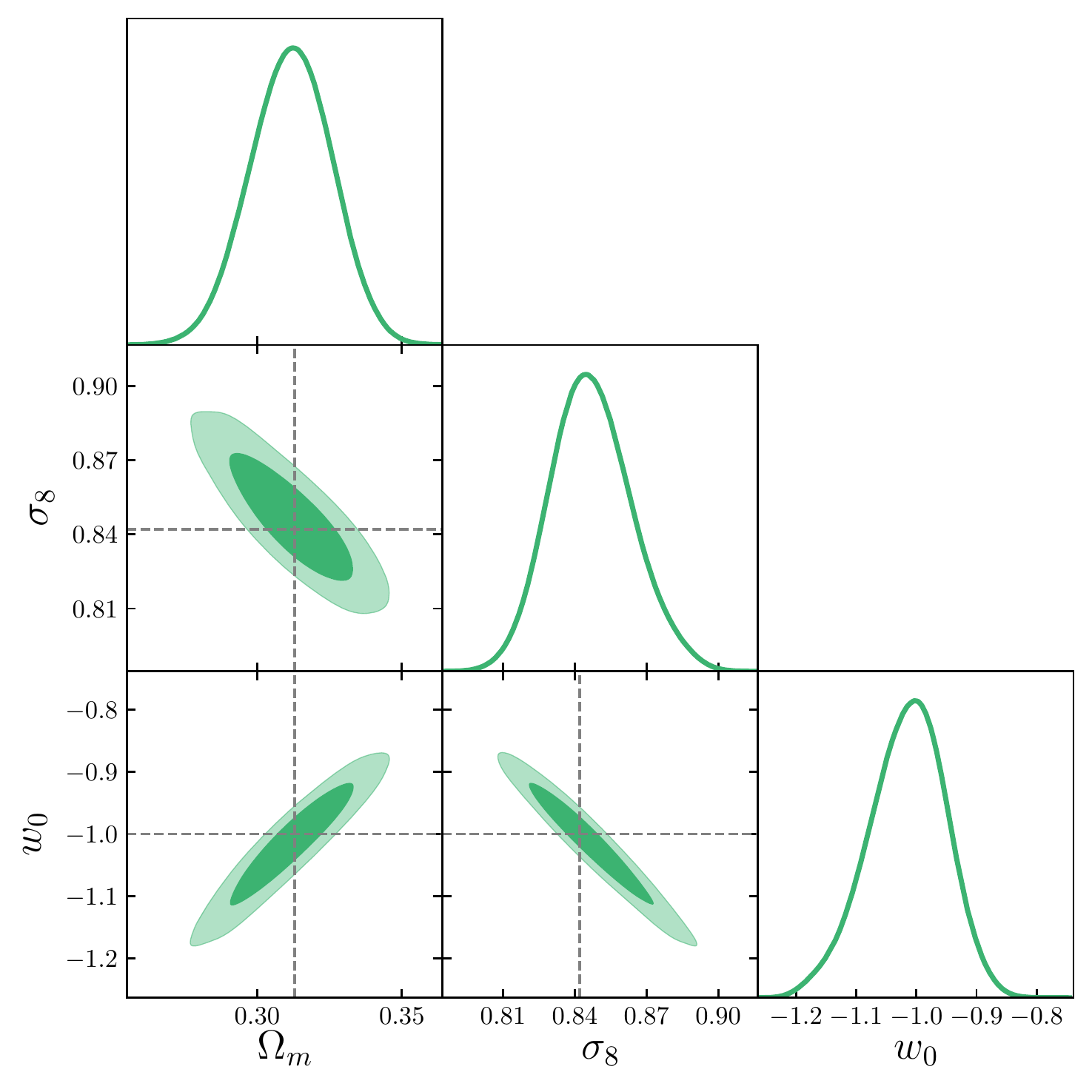}
    \caption{Posterior distribution of the main cosmological parameters for the shallow survey + BBN + BAO + CMB with a fiducial model generated using $f_{R0} = 10^{-4}$. All parameters are fully consistent with a vanilla $\Lambda$CDM model and none (including nuisance parameters not plotted here) show significant deviations $> 1 \sigma$ from their fiducial values indicated by dashed lines.}
    \label{fig:sdss_cmb_fR4fid_posterior}
\end{figure}

We find that the best-fit $w$CDM model does not show any significant deviations from the vanilla case. The full posterior distribution of the major cosmological parameters is shown in Fig.~\ref{fig:sdss_cmb_fR4fid_posterior}, and while there are small deviations in the nuisance parameters, all of them are within $1 \sigma$ compatible with their standard values without any peculiar features.

We also compare the richness distribution of cluster counts for the best-fit model with the $f(R)$ mock data in Fig.~\ref{fig:sdss_fR4_fid_counts} and find no significant deviations. The full parameter space (including the nuisance parameters described above) proves to be flexible enough to account even for a large value of $f_{R0}$ that could be detected if the correct model is assumed in the analysis. This indicates that $w$CDM might not be a good approach to search for generic deviations from $\Lambda$CDM for models that are not captured by this particular parametrisation.

Therefore we can not necessarily exclude modified gravity (or other) models just from the lack of tensions in the $\Lambda$CDM or $w$CDM analysis of cosmological surveys. Instead, it is necessary to consider the phenomenology of models individually in order to exclude them.

\begin{figure}
	\includegraphics[width=\columnwidth]{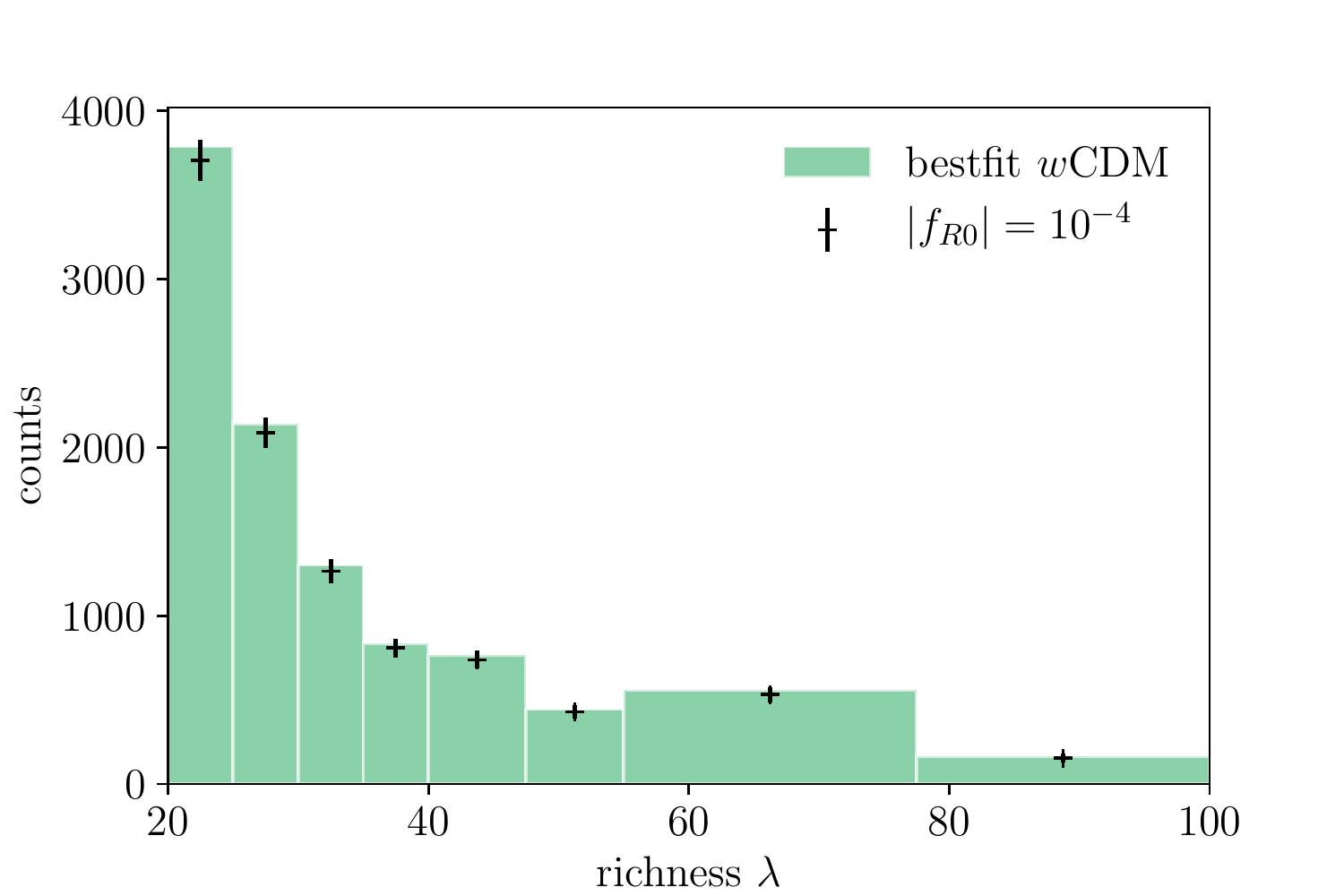}
    \caption{Bars show the binned richness distribution of clusters for the best-fit $w$CDM model compared to the fiducial data points generated with $f_{R0} = 10^{-4}$. All parameters for the best-fit model agree within $1 \sigma$ with their fiducial values.}
    \label{fig:sdss_fR4_fid_counts}
\end{figure}

\section{Conclusions}
\label{sec:conclusions}

In this paper, we presented an accurate halo mass function based on a spherical collapse framework valid for modified gravity and neutrino cosmologies, and calibrated it to a suite of specifically-designed cosmological simulations, the {\small DUSTGRAIN}-{\em pathfinder} runs. This allows joint constraints from cluster abundance studies. We keep the additional relative change and the fiducial GR mass function separate, so our results can be used with any other mass function calibrated to our mass definition of $M_{200m}$.

The cluster mass definition is crucial to accurately predict the characteristic $f(R)$ peak in the relative abundance because it governs the onset of screening effects. Mass functions for other commonly used mass definitions such as $M_{500c}$ therefore require recalibration of the screening mass, which we refer to future work.

We also demonstrate that the inclusion of neutrinos via a rescaling of the density field Eq.~\ref{eq:sigma_cdm} still holds in extended models, and we find a degeneracy between effects of $f(R)$ and massive neutrinos in the abundance of clusters that limits the ability of surveys with small redshift reach to disentangle them. This is likely to weaken existing limits on $f_{R0}$ from cluster abundance, and we will use the mass function for joint constraints using cluster data in a follow-up paper.

Deeper cluster surveys however can tell neutrinos and modified gravity reliably apart by their different redshift evolution, and future optical cluster samples will be able to probe the entire phenomenologically relevant parameter range of the model even when accounting for systematic uncertainties. This could be realised by the complete Dark Energy Survey, eRosita or Euclid\footnote{https://www.euclid-ec.org/} cluster samples.

We also explore the possibility to include $f(R)$ effects in the selection function of SZ or X-ray surveys directly as proposed by \cite{Schmidt_2010} and we find potentially large effects if the sample can be extended to include nearby, intermediate and low mass objects with $M \lesssim 10^{14} M_\odot/h$. Even though neutrinos can mask the additional abundance in the mass function at low redshifts, it is still possible to detect fifth forces through these selection effects. This allows to incorporate the limits on $f_{R0}$ from comparing lensing mass estimates and X-ray, SZ or dynamical mass estimates consistently into cluster abundance studies in a fully consistent framework.

Finally we find that generic searches for $w$CDM do not necessarily lead to significant tensions or conspicuous features when used to analyse mock $f(R)$ data -- even if the value of $f_{R0}$ could be detected with the same data set in a dedicated analysis. This emphasizes the need to model phenomenology of $\Lambda$CDM extensions carefully. A lack of tensions within a parametrisation does not imply the absence of new physics.

\section*{Acknowledgements}

Most cosmological quantities in this paper were calculated using the Einstein-Boltzmann code \texttt{CLASS} \citep{CLASS}.

SH wants to thank Vanessa B\"ohm and Korbinian Huber for many helpful discussions. We appreciate the help of Ben Moster with cross-checks for our simulation suite. JW and SH acknowledge the support of the DFG Cluster of Excellence "Origin and Structure of the Universe" and the Transregio programme TR33 "The Dark Universe".
MB acknowledges support from the Italian Ministry for Education,
University and Research (MIUR) through the SIR individual grant
SIMCODE (project number RBSI14P4IH),
from the grant MIUR PRIN 2015 "Cosmology and Fundamental Physics:
illuminating the Dark Universe with Euclid", and from the agreement ASI n.I/023/12/0 ``Attivit\`a relative alla fase B2/C per la missione Euclid".
The {\small DUSTGRAIN}-{\em pathfinder} simulations discussed in this work have been performed and analysed on the Marconi supercomputing machine at Cineca thanks to the PRACE project SIMCODE1 (grant nr. 2016153604) and on the computing facilities of the Computational Centre for Particle and Astrophysics (C2PAP) and the Leibniz Supercomputing Centre (LRZ) under the project ID pr94ji.

%%%%%%%%%%%%%%%%%%%%%%%%%%%%%%%%%%%%%%%%%%%%%%%%%%


\bibliographystyle{mnras}
\bibliography{Bibliography} % if your bibtex file is called example.bib

%%%%%%%%%%%%%%%%%%%%%%%%%%%%%%%%%%%%%%%%%%%%%%%%%%

% Don't change these lines
\bsp	% typesetting comment
\label{lastpage}
\end{document}